\documentclass[12pt]{article}
\usepackage[margin=2.5cm]{geometry}

\usepackage{makeidx}         
\usepackage{graphicx}        
\usepackage[bottom]{footmisc}

\usepackage{graphicx}
\usepackage{amsmath}
\usepackage{amssymb}
\usepackage{latexsym}
\usepackage{amsfonts}
\usepackage{slashed}
\usepackage{color}
\usepackage{bbold}

\usepackage{hyperref}

\def\hhref#1{\href{http://arxiv.org/abs/#1}{arXiv:#1}} 

\newcommand{\nn}{\nonumber\\}

\newcommand{\mc}{\mathcal}
\newcommand{\del}{\partial}
\newcommand{\bea}{\begin{eqnarray}}
\newcommand{\ea}{\end{eqnarray}}
\newcommand{\eea}{\end{eqnarray}}

\makeindex      

\begin{document}

\begin{titlepage}
\vfill
\vfill
\begin{center}
{\Large\bf The Chiral Magnetic Effect and Axial Anomalies\footnote{to appear in Lect. Notes Phys. "Strongly interacting matter in magnetic
fields" (Springer), edited by D. Kharzeev, K. Landsteiner, A. Schmitt, H.-U. Yee}}

\vskip 1in

G\"ok\c ce Ba\c sar$^*$\footnote{e-mail: {\tt basar@tonic.physics.sunysb.edu  }} and Gerald V. Dunne$^\dagger$\footnote{e-mail:
{\tt dunne@phys.uconn.edu}}
\vskip0.3in
 {$*$\it Department of Physics and Astronomy, Stony Brook University,} \\
{\it Stony Brook, NY 11794, USA }\\[0.3in]

{$^\dagger$\it Department of Physics, University of Connecticut,} \\
{\it Storrs, CT 06269, USA }\\[0.3in]

\end{center}

\vfill

\begin{abstract}
We give an elementary derivation of the chiral magnetic effect based on a strong magnetic field lowest-Landau-level projection in conjunction with the well-known axial anomalies in two- and four-dimensional space-time. The argument is general, based on a Schur decomposition of the Dirac operator. In the  dimensionally reduced theory, the chiral magnetic effect is directly related to the relativistic form of the Peierls instability, leading to a spiral form of the condensate, the chiral magnetic spiral. We then discuss the competition between spin projection, due to a strong magnetic field, and chirality projection, due to an instanton, for light fermions in QCD and QED. The resulting asymmetric distortion of the zero modes and near-zero modes is another aspect of the chiral magnetic effect.
\end{abstract}

\vfill

\end{titlepage}

\section{Dirac Operators, Dimensional Reduction and Axial Anomalies}
\label{sec:1}
In this first section we present an elementary relation between the chiral magnetic effect \cite{Kharzeev:2004ey,arXiv:0706.1026,arXiv:0711.0950,Fukushima:2008xe,Kharzeev:2009fn,Warringa:2012bq} and the axial anomalies in four-dimensional and two-dimensional space-time \cite{Jackiw:1983nv,Shifman:1999mk}. This follows from the basic structure of the Dirac operator, together with the lowest-Landau level (LLL) projection produced by a strong magnetic field, a basic feature of the phenomenon of magnetic catalysis \cite{Gusynin:1995nb}. We first review the sub-block structure of the Dirac operator and the associated  Schur decomposition \cite{strang}  of the propagator.

\subsection{Lowest Landau Level Projection}
\label{subsec:1.1}

We adopt the following conventions for Dirac matrices in four dimensional Minkowski space:
\begin{eqnarray}
\gamma^0=
\begin{pmatrix}
0 & {\mathbb 1} \cr
{\mathbb 1} & 0 
\end{pmatrix}
\quad, \quad
\gamma^j=
\begin{pmatrix}
0 & -\sigma^j \cr
\sigma^j & 0 
\end{pmatrix}
\quad, \quad
\gamma_5=
\begin{pmatrix}
  {\mathbb 1} & 0\cr
0& -{\mathbb 1}  
\end{pmatrix}
\quad, 
\label{dirac-4-m}
\end{eqnarray}
where $\sigma^j$, $j=1, 2, 3$, are the $2\times 2$ Pauli matrices, and ${\mathbb 1}$ is the $2\times 2$ identity matrix. It is convenient to write these $4\times 4$ Dirac matrices in $2\times 2$ block form:
\begin{eqnarray}
\gamma^\mu=
\begin{pmatrix}
0 & \alpha^\mu \cr
\tilde{\alpha}^\mu & 0 
\end{pmatrix}
\label{dirac-alpha}
\end{eqnarray}
where the $2\times 2$ matrices $\alpha^\mu$ and $\tilde{\alpha}^\mu$ are:
\begin{eqnarray}
\alpha^\mu=({\mathbb 1}, -\sigma^j)\qquad, \qquad \tilde{\alpha}^\mu=({\mathbb 1}, \sigma^j)
\label{alphas}
\end{eqnarray}
The Dirac matrices satisfy the anti-commutation relations, $\{ \gamma^\mu, \gamma^\nu\}=2\, \eta^{\mu\nu}$, with Minkowski metric $\eta^{\mu\nu}={\rm diag}(1, -1, -1, -1)$.

The Dirac operator is defined to be $D \hskip -6pt /=\gamma^\mu\left(\partial_\mu-i e A_\mu\right)$, where later we will take the gauge field $A_\mu$ to have both abelian and non-abelian parts, but for now we take it to be abelian. Thus, we have a natural  decomposition of the $4\times 4$ Dirac operator into $2\times 2$ sub-blocks:
\begin{eqnarray}
{\mathcal D}= -i \, D \hskip -6pt /+m &=&\begin{pmatrix}
m && -i \alpha^\mu D_\mu\cr
-i \tilde{\alpha}^\mu D_\mu && m\end{pmatrix} \nonumber\\
&\equiv & 
\begin{pmatrix}
m && D\cr
-\tilde{D}  && m\end{pmatrix}
\label{dirac-decomp}
\end{eqnarray}
Since $\det\left(-i \, D \hskip -6pt /+m\right)=\det\left(i \, D \hskip -6pt /+m\right)$, we consider 
\begin{eqnarray}
\det\left(i \, D \hskip -6pt /+m\right)\det\left(-i \, D \hskip -6pt /+m\right)=
\begin{pmatrix}
m^2+D\, \tilde{D} && 0 \cr
0 && m^2+\tilde{D}\, D
\end{pmatrix}
\label{dsquare}
\end{eqnarray}
The operators $D\, \tilde{D}$ and $\tilde{D}\, D$ differ from one another, and from the Klein-Gordon operator $D_\mu D^\mu$,  in their spin-projection terms:
\begin{eqnarray}
D\, \tilde{D}&=& D_\mu D^\mu +\frac{e}{2}\tilde\sigma^{\mu\nu}F_{\mu\nu}
\label{ddt}\\
\tilde{D}\, D&=& D_\mu D^\mu +\frac{e}{2}{\sigma}^{\mu\nu}F_{\mu\nu}
\label{dtd}
\end{eqnarray}
where the $2\times 2$ spin matrices, 
\begin{eqnarray}
\tilde\sigma^{\mu\nu}=\frac{1}{2i}\left(\alpha^\mu \tilde{\alpha}^\nu-\alpha^\nu\tilde{\alpha}^\mu\right)\quad, \quad 
{\sigma}^{\mu\nu}=\frac{1}{2i}\left(\tilde{\alpha}^\mu \alpha^\nu-\tilde{\alpha}^\nu\alpha^\mu\right)
\label{sigmas}
\end{eqnarray}
are the sub-block components of the usual $4\times 4$ spin matrices 
\begin{eqnarray}
\Sigma^{\mu\nu}=\frac{1}{2i}[\gamma^\mu, \gamma^\nu]=
\begin{pmatrix}
\tilde\sigma^{\mu\nu} & 0\cr
0&{\sigma}^{\mu\nu}
\end{pmatrix}
\label{4spin}
\end{eqnarray}

Now, suppose we have a background electromagnetic field consisting of a magnetic field $\vec{B}=(0, 0, B)$, and an electric field $\vec{E}=(0, 0,  E)$, both of which are pointing in the $x^3$ direction. We do not need to assume that these fields are uniform, but we choose $B=B(x^1, x^2)$ and $E=E(x^0, x^3)$, choosing gauge field $A^\mu=(A^0(x^3), A^1(x^2), A^2(x^1), A^3(x^0))$, which satisfies $\partial_\mu A^\mu=0$. Then, with only a third component of both the electric and magnetic field,  the $2\times 2$ operators in (\ref{ddt}) and (\ref{dtd}) reduce to 
\begin{eqnarray}
D\, \tilde{D}&=& D_\mu D^\mu -e\left(B+i\, E\right)\sigma^3
\label{ddt2}
\\
\tilde{D}\, D&=& D_\mu D^\mu -e\left(B-i\,  E\right)\sigma^3
\label{dtd2}
\end{eqnarray}
because $\sigma^{12}=\tilde{\sigma}^{12}=-\sigma^3$, while $\sigma^{03}=-\tilde{\sigma}^{03}=i\sigma^3$.
The important point in (\ref{ddt2}) and (\ref{dtd2}) is that both $2\times 2$ operators $D\, \tilde{D}$ and $\tilde{D}\, D$ are {\bf diagonal} in terms of their Dirac matrix structure.

To understand the effect of a strong magnetic field, and in particular its projection to the lowest Landau level (LLL), consider the factorization of the Klein-Gordon operator, for this case parallel $\vec{E}$ and $\vec{B}$ in the $x^3$ direction:
\begin{eqnarray}
D_\mu D^\mu=(D_0\mp D_3)(D_0\pm D_3)\pm i\, e\, E -(D_1\mp i D_2)(D_1\pm i D_2) \pm e\, B
\label{bogom}
\end{eqnarray}
If the magnetic field is constant, and for example in the symmetric gauge $A_1=-\frac{B}{2}x^2$, and $A_2=\frac{B}{2}x^1$, then we can adopt complex coordinates, $z=x^1+i x^2$, $\bar{z}=x^1-i x^2$, to write
\begin{eqnarray}
D_1\pm i D_2=\begin{cases}
2\left(\partial_{\bar{z}}+\frac{e B}{4}\, z\right) \cr
2\left(\partial_{z}-\frac{e B}{4}\, \bar{z}\right) 
\end{cases}
\label{holo}
\end{eqnarray}
If $e B>0$ we choose the upper sign to obtain normalizable solutions, with the Gaussian factor $\exp(-\frac{eB}{4} |z|^2)$, in which case we see that the $e B$ term in (\ref{bogom}) cancels the spin term in (\ref{ddt2}) and (\ref{dtd2}), when $\sigma^3= +1$. Thus the magnetic component of the Dirac operators leads to a zero mode of the $2\times 2$ operators $D\, \tilde{D}$ and $\tilde{D}\, D$, when the spin is aligned along the direction of the magnetic field. In fact, this also applies to the situation of an inhomogeneous $B(x^1, x^2)$ field, and the degeneracy of this lowest-Landau-level (LLL) is given by the integer part of the net magnetic flux; this is the Aharonov-Casher theorem \cite{aharonov-casher,novikov},  the projection onto the LLL. 

\subsection{Schur decomposition of Dirac propagator}
\label{subsec:1.2}
The Schur decomposition gives an elementary algebraic decomposition of the inverse of a matrix in terms of its sub-block structure \cite{strang}. This leads immediately to an associated sub-block decomposition of the Dirac propagator, and as we show, it also provides a simple description of the lowest Landau level projection in a strong magnetic field, which is the key to magnetic catalysis \cite{Gusynin:1995nb}.

Consider a matrix $M$ written as
\begin{eqnarray}
M=\begin{pmatrix}
a & b\cr
c& d
\end{pmatrix}
\label{schur-m}
\end{eqnarray}
where $a$ and $d$ are square, but not necessarily of the same size [and so $b$ and $c$ need not be square matrices], then we can write the inverse of $M$  in block form in two different ways: 
\begin{eqnarray}
M^{-1}&=& 
\begin{pmatrix}
s^{-1} &&&& -s^{-1}\, b\, d^{-1} \cr 
-d^{-1}\, c\, s^{-1} &&&& d^{-1} \,c\, s^{-1}\, b\, d^{-1}+d^{-1}
\end{pmatrix} \quad, \quad s\equiv a-b\, d^{-1}\, c\\
&=& \begin{pmatrix}
a^{-1} \,b\, t^{-1}\, c\, a^{-1}+a^{-1}&&&& -a^{-1}\, b\, t^{-1} \cr 
-t^{-1}\, c\, a^{-1} &&&& t^{-1}
\end{pmatrix} \quad, \quad t\equiv d-c\, a^{-1}\, b
\label{schur}
\end{eqnarray}
where $s$ and $t$ are the two different Schur complements of $M$. Note that the first expression only requires $d$ and $s$ to be invertible, while the second expression only requires $a$ and $t$ to be invertible. Applying this to the Dirac operator in (\ref{dirac-decomp}), we find 
\begin{eqnarray}
s=\frac{1}{m}\left(m^2+D\, \tilde{D}\right)\qquad, \qquad t=\frac{1}{m}\left(m^2+ \tilde{D}\, D\right)
\label{st}
\end{eqnarray}
which gives then two different decompositions of the Dirac propagator:
\begin{eqnarray}
{\mathcal D}^{-1}&=& 
\begin{pmatrix}  \frac{m}{m^2+D\, \tilde{D}} &&&&  \frac{1}{m^2+D\, \tilde{D}} \, D \cr 
\tilde{D}  \frac{1}{m^2+D\, \tilde{D}}  &&&& \frac{1}{m}\left(1-\tilde{D}  \frac{1}{m^2+D\, \tilde{D}} \, D\right)\end{pmatrix}
\label{s-prop1}
\\
&=&
\begin{pmatrix}  
 \frac{1}{m}\left(1-\tilde{D}  \frac{1}{m^2+D\, \tilde{D}} \, D\right) &&&&  D\frac{1}{m^2+\tilde{D}\, D} \cr 
 \frac{1}{m^2+ \tilde{D}\, D}\, \tilde{D}  &&&& \frac{m}{m^2+D\, \tilde{D}} \end{pmatrix}
\label{s-prop2}
\end{eqnarray}
The Euclidean analogue of this decomposition corresponds precisely to the chiral decompositions used in \cite{Brown:1977bj,Hur:2010bd}:
\begin{eqnarray}
{\mathcal D}^{-1}&=& 
{\mathcal D}\, {\mathcal G}^{(\pm)}\left( \frac{{\mathbb 1}\pm \gamma_5}{2}\right)+
{\mathcal G}^{(\pm)}\, D\hskip -6pt /\left( \frac{{\mathbb 1}\mp \gamma_5}{2}\right)
+\frac{1}{m}\left(1-D\hskip -6pt /\,  {\mathcal G}^{(\pm)} D\hskip -6pt /\right) \left( \frac{{\mathbb 1}\mp \gamma_5}{2}\right)\nn
\label{chiral-prop}
\end{eqnarray}
where ${\mathcal G}^{(+)}=1/(m^2+D\, \tilde{D})$, and ${\mathcal G}^{(-)}=1/(m^2+ \tilde{D}\, D)$. These chiral  decompositions of the fermion propagator are particularly useful since they show clearly the projection onto chiral zero modes.

In order to use such Schur decompositions to characterize also the lowest Landau level projection, in addition to the chiral decomposition,  it is convenient to write the propagators in (\ref{s-prop1}) and (\ref{s-prop2}) in factored form:
\begin{eqnarray}
{\mathcal D}^{-1}&=& \frac{1}{m}
\begin{pmatrix} {\mathbb 1} && 0\cr \frac{1}{m}\tilde{D}&& {\mathbb 1}\end{pmatrix}
\begin{pmatrix}  \frac{m^2}{m^2+D\, \tilde{D}} && 0\cr 0 && {\mathbb 1}\end{pmatrix}
\begin{pmatrix} {\mathbb 1} && - \frac{1}{m} D\cr 0 && {\mathbb 1}\end{pmatrix}
\label{schur-prop1}
\\
&=& \frac{1}{m}
\begin{pmatrix} {\mathbb 1} && \frac{1}{m} D\cr 0 && {\mathbb 1}\end{pmatrix}
\begin{pmatrix} {\mathbb 1} && 0\cr 0 && \frac{m^2}{m^2+\tilde{D}\, D}\end{pmatrix}
\begin{pmatrix} {\mathbb 1} && 0\cr -\frac{1}{m}\tilde{D}&& {\mathbb 1}\end{pmatrix}
\label{schur-prop2}
\end{eqnarray}
We stress that at this point we have only used elementary algebra to express a general Dirac propagator in terms of  its $2\times 2$ sub-block structure.

From these Schur decompositions, we can reduce the expectation values of the charge and axial currents to much simpler forms. By straightforward manipulations we find for the charge current
\begin{eqnarray}
< j^\mu> &=& i\, e\, {\rm tr}_{4\times 4}\left(\gamma^\mu \, {\mathcal D}^{-1}\right)\\
 &=& i\, e\, {\rm tr}_{2\times 2}\left(\left(D\, \tilde{\alpha}^\mu-\alpha^\mu \tilde{D}\right)\frac{1}{m^2+D\, \tilde{D}}\right)\\
  &=& 2 e\,\eta^{\mu\nu} {\rm tr}_{2\times 2}\left(D_\nu \frac{1}{m^2+D\, \tilde{D}}\right)
\label{jmu}
\end{eqnarray}
and for the axial current
\begin{eqnarray}
< j^\mu_5> &=& i\, e\, {\rm tr}\left(\gamma^\mu\, \gamma_5 \, {\mathcal D}^{-1}\right)\\
 &=& i\, e\, {\rm tr}_{2\times 2}\left(\left(D\, \tilde{\alpha}^\mu+\alpha^\mu \tilde{D}\right)\frac{1}{m^2+D\, \tilde{D}}\right)\\
  &=& 2 i\, e\, {\rm tr}_{2\times 2}\left(\tilde\sigma^{\mu\nu}\, D_\nu \frac{1}{m^2+D\, \tilde{D}}\right)
\label{j5mu}
\end{eqnarray}
Here we have used the facts that $(\alpha^\nu\,\tilde{\alpha}^\mu+\alpha^\mu\, \tilde{\alpha}^\nu)=2\eta^{\mu\nu}\, {\mathbb 1}_{2\times 2}$, while $(\alpha^\nu\,\tilde{\alpha}^\mu-\alpha^\mu\, \tilde{\alpha}^\nu)=2i\tilde\sigma^{\mu\nu}$. In particular, note that since $\tilde\sigma^{03}=-i\, \sigma^3=-\tilde\sigma^{30}$,
\begin{eqnarray}
< j^0> &=& 2 e\,\eta^{\mu\nu} {\rm tr}_{2\times 2}\left(D_0 \frac{1}{m^2+D\, \tilde{D}}\right) \\
< j^3> &=&- 2 e\,\eta^{\mu\nu} {\rm tr}_{2\times 2}\left(D_3 \frac{1}{m^2+D\, \tilde{D}}\right)  \\
< j^0_5> &=& 2 e\,\eta^{\mu\nu} {\rm tr}_{2\times 2}\left(\sigma^3\, D_3 \frac{1}{m^2+D\, \tilde{D}}\right) \\
< j^3_5> &=&- 2 e\,\eta^{\mu\nu} {\rm tr}_{2\times 2}\left(\sigma^3\, D_0 \frac{1}{m^2+D\, \tilde{D}}\right)
\label{03}
\end{eqnarray}
Again, thus far we have only used elementary algebra to reduce the expectation values of the $4\times 4$ matrices and propagators to expressions involving just $2\times 2$ matrices and propagators.

\subsection{Currents and anomalies in the lowest Landau level projection}
\label{subsec:1.3}

Now suppose the background field consists of a very strong magnetic field in the $x^3$ direction. Then we project onto the LLL, which means that the states contributing to the $2\times 2$ expectation values have $\sigma^3=+1$. Therefore, we see immediately that in this LLL projection limit, where we project onto motion along the magnetic field direction, we can write for the remaining currents:
\begin{eqnarray}
< j^M_5> =\epsilon^{MN} < j_N>\qquad, \qquad M, N =0, 3
\label{22}
\end{eqnarray}
where the epsilon symbol is $\epsilon^{03}=+1=-\epsilon^{30}$. This is exactly the relation between the charge and axial current in two dimensional space-time.

Furthermore, suppose the four dimensional background field consists of a strong magnetic field and also an electric field, both directed along the $x^3$ axis. Then, following the analysis of the first section, we choose a gauge field of the form $A^\mu=(A^0(x^3), A^1(x^2), A^2(x^1), A^3(x^0))$,  satisfying $\partial_\mu A^\mu=0$. Then a simple two dimensional computation yields
current expectation values for Dirac indices $0$ and $3$ (we adopt the convention of using capital Roman indices $M, N$ to denote the components of the dimensionally reduced $(x^0, x^3)$ plane):
\begin{eqnarray}
< j^M> =\frac{e B}{2\pi}\, \frac{e A^M}{\pi} \qquad, \qquad 
< j^M_5> =\frac{e B}{2\pi}\, \epsilon^{MN}\frac{e A_N}{\pi}
\label{2dj}
\end{eqnarray}
where $\frac{e B}{2\pi}$ is the Landau degeneracy factor, in its local Aharonov-Casher form.
Note that these expressions are consistent with charge current conservation and the two dimensional axial anomaly:
\begin{eqnarray}
\partial_M < j^M> =0 \qquad, \qquad \partial_M < j^M_5> = \frac{e B}{2\pi}\, \frac{e E}{\pi} 
\label{2d-anomaly}
\end{eqnarray}

In this LLL projection limit, we can alternatively express this result in four dimensional language as
\begin{eqnarray}
\partial_\mu < j^\mu>_{LLL} =0 \qquad, \qquad \partial_\mu < j^\mu_5>_{LLL} = \frac{e^2}{2\pi^2}\vec{B}\cdot\vec{E}=\frac{e^2}{8\pi^2} F^{\mu\nu}\tilde{F}_{\mu\nu}  
\label{4d-anomaly}
\end{eqnarray}
which expresses  charge current conservation and the four dimensional axial anomaly. This makes it clear that the relevant anomaly is the ``covariant'' anomaly, rather than the ``consistent'' anomaly. For abelian theories these differ by a factor of $1/d$ in $(2d-2)$ space-time dimensions, while for non-abelian theories the covariant and consistent anomalies have different field structure \cite{bardeen,dt}.

To make the connection with the chiral magnetic effect, we note that it is natural to identify $A^0(x^3)$ with a spatially dependent chemical potential, and $A^3(x^0)$ with a time dependent chiral chemical potential:
\begin{eqnarray}
A^0\quad \leftrightarrow\quad \mu\qquad, \qquad A^3\quad\leftrightarrow\quad \mu_5
\label{ident}
\end{eqnarray} 
For $A^0$ and $\mu$ this is obvious, 
because the coupling is given by $\mu\, \bar{\psi}\gamma^0 \psi$. For $A^3$ and $\mu_5$, this follows because the coupling is $\mu_5\, \bar\psi \gamma_5\gamma^0\psi$, and in the LLL projection $\gamma_5\gamma^0\leftrightarrow \gamma^3$, since $\gamma^3$ has off-diagonal sub-blocks $\mp \sigma^3$, and $\sigma^3\to +1$ in the LLL limit.

Therefore, we can understand the two dimensional currents in (\ref{2dj}) as
\begin{eqnarray}
< j^0> =\frac{\mu}{\pi}\, \frac{e B}{2\pi} \qquad, \qquad < j^3> =\frac{\mu_5}{\pi}\, \frac{e B}{2\pi}\\
< j^0_5> =\frac{\mu_5}{\pi}\, \frac{e B}{2\pi}  \qquad, \qquad < j^3_5> =\frac{\mu}{\pi}\, \frac{e B}{2\pi}
\label{cme}
\end{eqnarray}
These relations express the chiral magnetic effect, which we see is a direct consequence of the axial anomalies in two and four dimensional space-time, after the LLL projection caused by a strong magnetic field. The coefficients are fixed completely by the anomaly equations. For a complementary discussion of the relation between chiral asymmetry and the axial anomaly, see  \cite{Gorbar:2010kc}.
\begin{figure}[htb]
\includegraphics[scale=0.25]{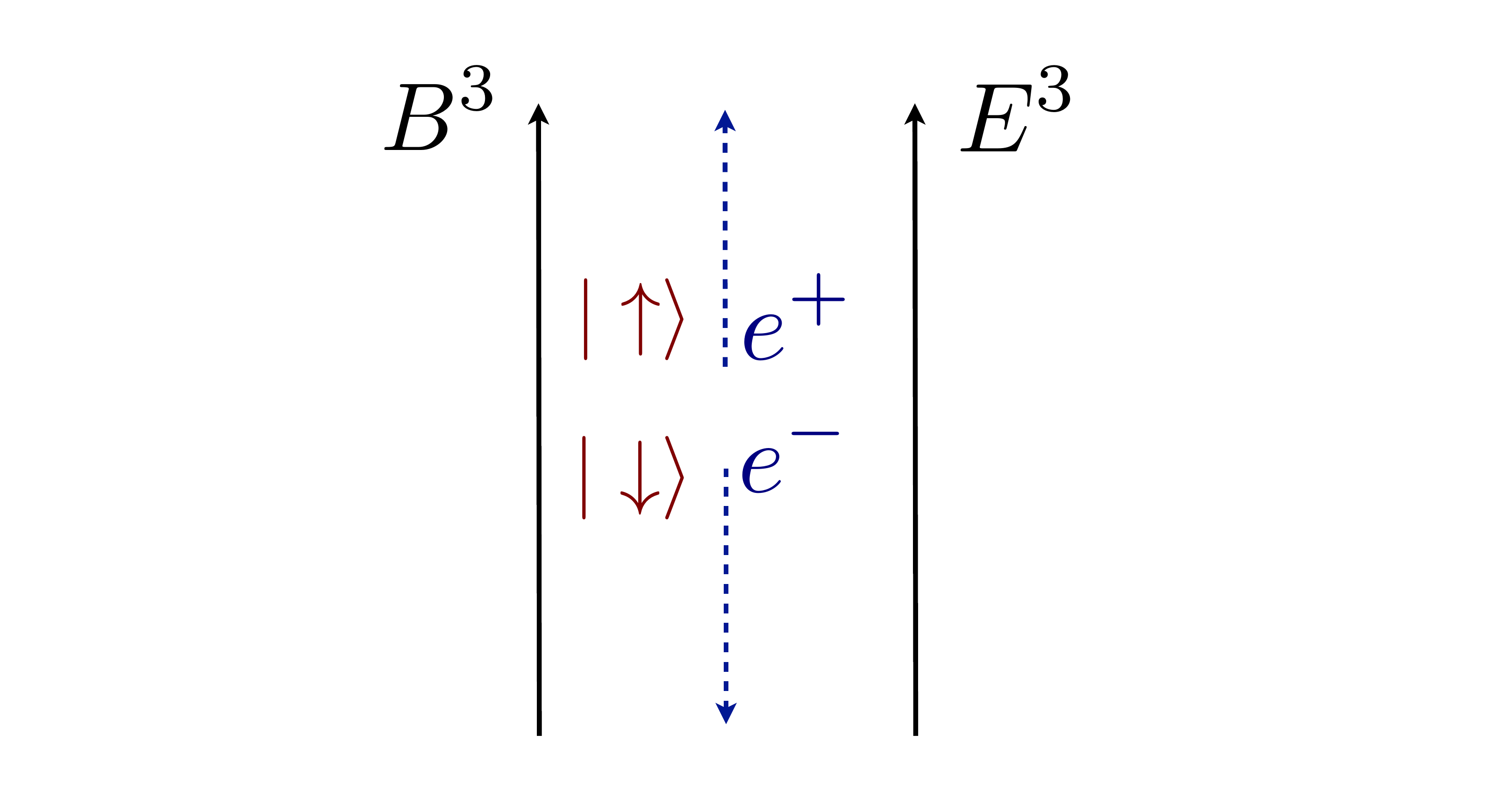}
\caption{Parallel electric and magnetic fields produce electron-positron pairs from vacuum, and because of the LLL projection caused by the strong magnetic field, correlated with spin and charge, this results in a net flow of chirality along the direction of the fields, in accordance with the Schwinger pair production rate and the chiral magnetic effect.}
\label{fig1}
\end{figure}

\subsection{Chiral Magnetic Effect and the Schwinger Effect}
\label{sec:1.4}

The chiral magnetic effect can also be understood naturally in terms of the Schwinger effect \cite{he,schwinger1}, particle production from vacuum, which occurs when there is a non-zero electric field background, as illustrated in Figure \ref{fig1}. 
With approximately constant parallel electric and magnetic fields directed along the $x^3$ axis, the Schwinger pair production rate, per unit volume,  is given by
\begin{eqnarray}
\Gamma = e^2 \frac{E\, B}{4\pi^2} \coth\left(\frac{B}{E}\,\pi\right)\, e^{-m^2 \pi/|e E|}
\label{schwinger}
\end{eqnarray}
When $B\to 0$ we recover the usual result for a pure electric field, and for nonzero magnetic field we find an enhancement of the rate which is linear in $B$ in the strong magnetic field limit. Positrons are accelerated along the direction of the electric field, and electrons in the opposite direction. 
In the massless fermion limit this corresponds to a net production and flow of chirality, because the lowest-Landau-level projection projects spin according to the charge and the direction of the magnetic field. 
Thus we find the rate of change of chirality
\begin{eqnarray}
\frac{d j^0_5}{dt} = 2\, \Gamma =\frac{E\, B}{2\pi^2}
\label{schwinger-cme}
\end{eqnarray}
in agreement with the axial anomaly (\ref{4d-anomaly}) and the chiral magnetic effect (\ref{cme}). The electric field produces the acceleration while the strong magnetic field provides the LLL projection that correlates spin with the direction of flow of charge, and hence also of chirality. Physically, a spatially dependent $A^0(x^3)$ produces charge separation, as for a local chemical potential, while a  time dependent $A^3(x^0)$ drives a current along the direction of the electric field \cite{Kluger:1998bm}, which in the LLL projection corresponds to a flow of chirality.

\subsection{Maxwell-Chern-Simons theory and the Schwinger model}
\label{sec:1.5}

The interpretation of the chiral magnetic effect in terms of the effect of an electric field directed along the same direction as the strong magnetic field is also very natural in terms of an effective Maxwell-Chern-Simons theory resulting from an adiabatic space- or time-dependent theta parameter \cite{Kharzeev:2009fn}. Express the theta term in the Lagrangian as (up to a total derivative)
\begin{eqnarray}
-\frac{e^2}{8 \pi^2}\,\theta\, F^{\mu\nu}\tilde{F}^{\mu\nu} =P_\mu J^\mu_{CS}
\label{mcs}
\end{eqnarray}
where
\begin{eqnarray}
P_\mu=\partial_\mu \theta\qquad, \qquad J^\mu_{CS}=\frac{e^2}{8 \pi^2}\,\epsilon^{\mu\nu\rho\sigma}A_\nu F_{\rho\sigma}
\label{theta}
\end{eqnarray}
The pseudo vector $P_\mu$ encodes the anomalous terms from the chiral magnetic effect, modifying the usual inhomogeneous Maxwell equations to read
\begin{eqnarray}
\partial_\mu F^{\mu\nu}=J^\nu-\frac{e^2}{2\pi^2}\,P_\mu \tilde{F}^{\mu\nu}
\label{mod}
\end{eqnarray}
\begin{figure}[htb]
\center\includegraphics[scale=0.27]{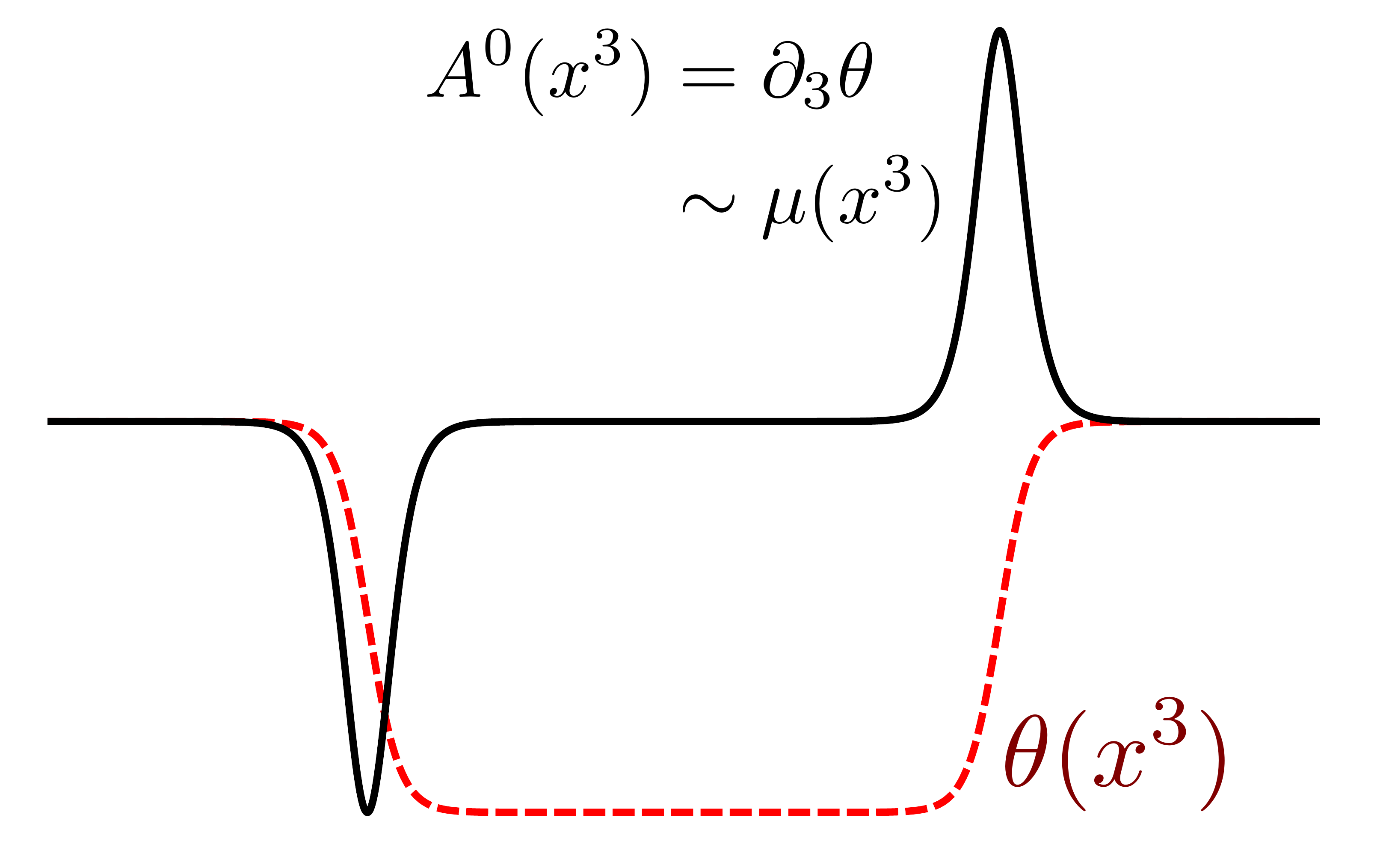}\\
\includegraphics[scale=0.27]{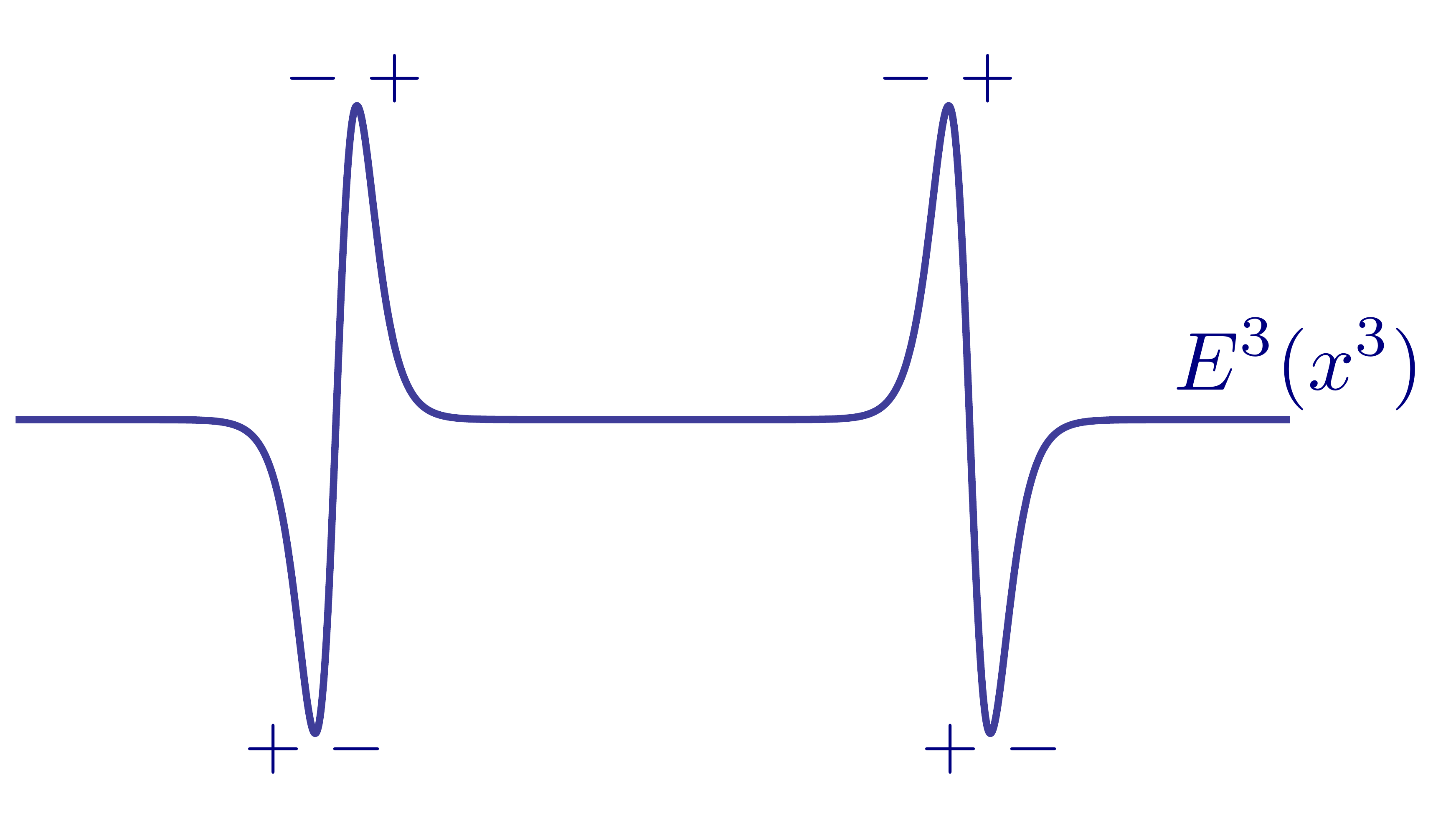}
\caption{The effect of a spatially inhomogeneous theta parameter, with the interpretation of its gradient being a spatially inhomogeneous chemical potential $\mu\sim A^0$. The resulting spatially inhomogeneous electric field produces a build up of negative charge at the left-hand inhomogeneity and positive charge at the right-hand inhomogeneity, producing the electric charge separation of the chiral magnetic effect.}
\label{fig2}
\end{figure}
\begin{figure}[htb]
\center\includegraphics[scale=0.27]{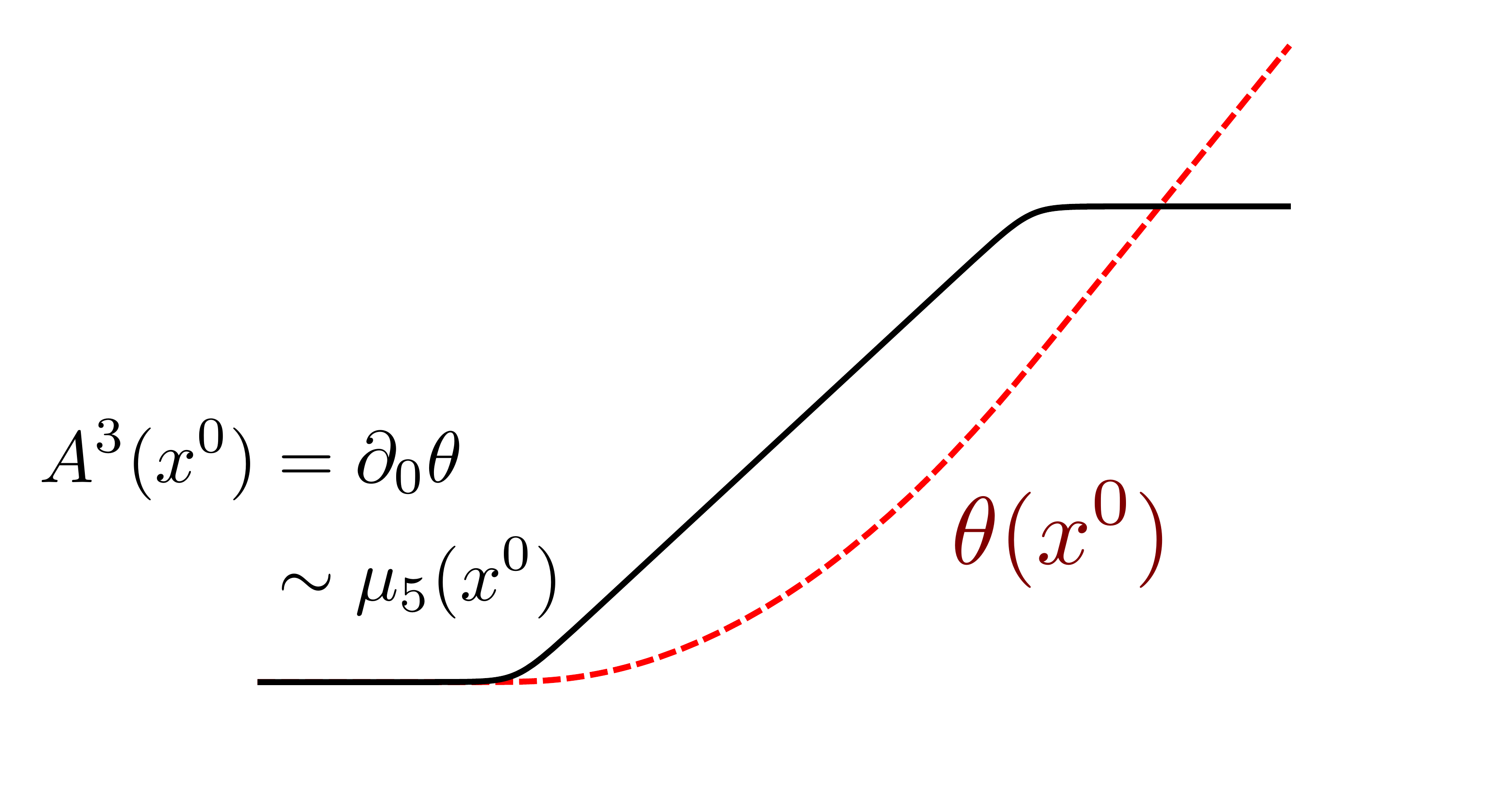}\\
\includegraphics[scale=0.27]{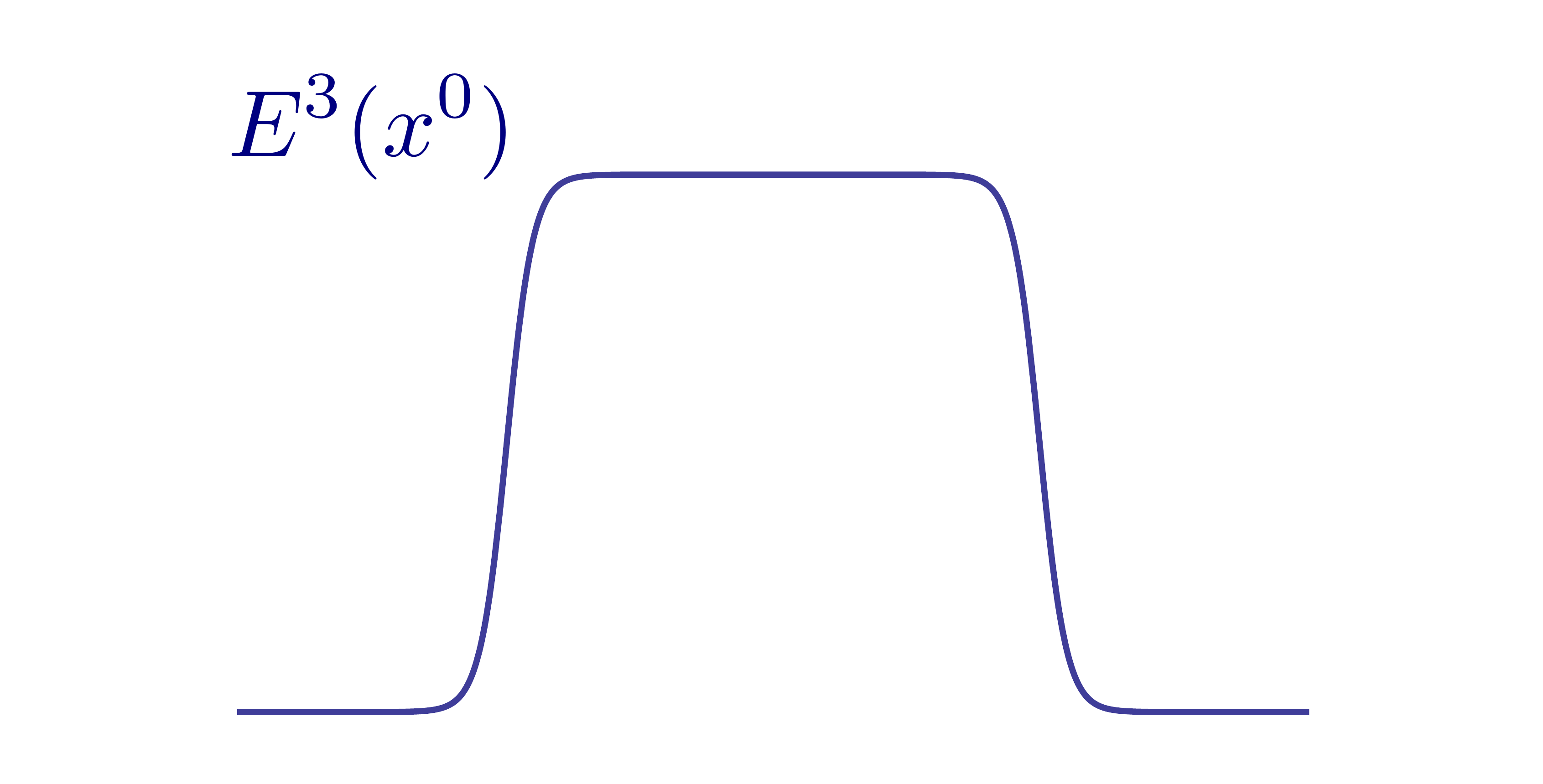}
\caption{The effect of a time dependent theta parameter, with the interpretation of its gradient being a time-dependent chiral chemical potential $\mu_5\sim A^3$ in the LLL projection. Via the Schwinger effect, the resulting time dependent  electric field produces a flow of electric current  linear in $A^3(t)$, and in the LLL projection this corresponds to a net flow of chirality.}
\label{fig3}
\end{figure}

From (\ref{cme}), in the strong magnetic field limit  we project to the 2d $(x^0, x^3)$ plane with the identifications
\begin{eqnarray}
P_0=A_3\quad, \quad P_3=A_0\quad \Rightarrow \qquad P^M=\epsilon^{MN}A_N \quad
\label{pmu}
\end{eqnarray}
Furthermore, in the strong magnetic field limit, the Chern-Simons current projects to the 2d $(x^0, x^3)$ plane as
\begin{eqnarray}
J_{CS}^M=- \frac{e^2\,B}{4 \pi^2}\, \epsilon^{MN}A_N
\label{cs}
\end{eqnarray}
Therefore, in the strong magnetic field limit, the theta term in (\ref{mcs}) reduces to a mass term of the 2d gauge field, providing an explicit realization of the 2d Schwinger model \cite{Schwinger:1962tp,Jackiw:1983nv}, with the effective Maxwell equations in the reduced 2d $(x^0, x^3)$ plane being written as
\begin{eqnarray}
J^M =\left(\square +\frac{e^2 B}{2\pi^2}\right)A^M 
\label{sm}
\end{eqnarray}

In physical terms, we see from (\ref{pmu}) that a spatial inhomogeneity in the theta parameter corresponds to a non-zero $A^0(x^3)$, which is a spatially inhomogeneous chemical potential. We expect this to produce charge separation. In terms of the Schwinger effect, this is illustrated in Figure \ref{fig2}, depicting a spatial region of non-zero $\theta$. At the edges, the gradient is non-zero, which produces a spatially inhomogeneous electric field, as shown in the Figure. This leads to an accumulation of opposite charges at the edges of the region of non-zero theta.
On the other hand, a time-dependence in the theta parameter corresponds to a non-zero $A^3(x^0)$, which acts as a chiral chemical potential in the LLL projection limit. For example, consider an electric field turning on smoothly at some early time, and turning off smoothly at some later time, as shown in Figure \ref{fig3}. Then the $A^3(x^0)$ field is approximately constant over the time of nonzero electric field, and this is known to drive a current $< J^3>$ that is linear in $A^3(x^0)$ \cite{Kluger:1998bm}, in agreement with the chiral magnetic effect relation (\ref{cme}).

\section{Chiral Magnetic Spiral}
\label{sec_cms}
In this Section we  elaborate more on the dimensional reduction to two-dimensions due to strong magnetic field. In particular, we will show that the expectation values of vector and axial currents, $<J^\mu>$ and $<J_5^\mu>$, can be expressed in terms of various fermionic bilinears whose dynamics are governed by a two-dimensional Lagrangian. The two dimensional axial anomaly reproduces the well known chiral magnetic effect. Furthermore, in addition to the chiral magnetic effect, the universal dynamics of two-dimensional chiral fermions implies the existence of additional currents which are {\bf transverse} to the magnetic field, and have a spiral modulation  along the direction of the field: the ``chiral magnetic spiral'' \cite{Basar:2010zd}. 

\subsection{Basic setup and dimensional reduction}

We decompose the 4-component spinor in terms of eigenstates of the chiral projectors, $P_{R,L}=\frac{1}{2}({\bf 1}\pm \gamma^5)$, the spin projectors $P_{\uparrow,\downarrow}=\frac{1}{2}({\bf 1}\pm \Sigma^3)$, and the momentum direction projectors $P_{+,-}=\frac{1}{2}({\bf 1}\pm \gamma^0\gamma^3)$. 
The longitudinal spin operator is  
$\Sigma^3=\gamma^0\gamma^3\gamma^5 ={\rm diag}(
\sigma^3,\sigma^3)$, and   
$\gamma^0\gamma^3 ={\rm diag}(
\sigma^3,-\sigma^3)$.
We can write the 4-component  spinor field as
\begin{equation}
\Psi=\begin{pmatrix}
R_+\cr
R_-\cr
L_-\cr
L_+\cr
\end{pmatrix}
\end{equation}
The four-component spinor can be decomposed into two-component sub spinors in various ways. The chirality and spin decompositions respectively are
\bea
&\psi_R\,=\begin{pmatrix}
R_+\cr
R_-\cr
\end{pmatrix} \quad , \quad
&\psi_L\,=\begin{pmatrix}
L_+\cr
L_-\cr
\end{pmatrix} 
\nonumber \\
&\phi_\uparrow=\begin{pmatrix}
R_+\cr
L_-\cr
\end{pmatrix}\quad  , \quad
&\phi_\downarrow=\begin{pmatrix}
L_+\cr
R_-\cr
\end{pmatrix}
\ea
where $\pm$ denotes the direction of motion along $x_3$, the direction of the magnetic field.
The corresponding four-dimensional currents can as well be decomposed in terms of chirality and spin sub-spinors. The vector current $\bar\Psi\gamma^\mu\Psi$ has the decomposition
\bea
J^0&=&\psi_R^\dagger \psi_R+\psi_L^\dagger \psi_L
=\phi_\uparrow ^\dagger \phi_\uparrow+\phi_\downarrow^\dagger \phi_\downarrow
\nonumber \\
J^1&=& \bar\psi_R\,\psi_R-\bar\psi_L\,  \psi_L
=-\bar\phi_\uparrow \Gamma^5 \phi_\downarrow+ \bar\phi_\downarrow  \Gamma^5 \phi_\uparrow
\nonumber \\
J^2&=&i\bar\psi_R\,\Gamma^5 \psi_R+i\bar\psi_L\, \Gamma^5 \psi_L
=i \bar\phi_\uparrow \Gamma^5 \phi_\downarrow+ i\bar\phi_\downarrow \Gamma^5 \phi_\uparrow
\nonumber \\
J^3&=& \bar\psi_R\,\Gamma^z \psi_R+\bar\psi_L\,\Gamma^z \psi_L
=\bar{\phi_\uparrow} \Gamma^z\phi_\uparrow+\bar{\phi_\downarrow} \Gamma^z\phi_\downarrow
\label{vector}
\ea
The axial current $J_5^\mu=\bar{\Psi}\gamma^\mu\gamma^5\Psi$  has a similar form:
\bea
J_5^0&=&{\psi_R}^\dagger \psi_R-\psi_L^\dagger \psi_L
=-i \bar\phi_\uparrow \Gamma^z \phi_\uparrow+i \bar\phi_\downarrow \Gamma^z \phi_\downarrow
\nonumber \\
J_5^1&=& \bar\psi_R\,\psi_R+\bar\psi_L\,  \psi_L
=\bar\phi_\uparrow  \phi_\downarrow+ \bar\phi_\downarrow  \phi_\uparrow
\nonumber \\
J_5^2&=&i\bar\psi_R\,\Gamma^5 \psi_R-i\bar\psi_L\, \Gamma^5 \psi_L
=-i \bar\phi_\uparrow  \phi_\downarrow+ i\bar\phi_\downarrow  \phi_\uparrow
\nonumber \\
J_5^3&=& \bar\psi_R\,\Gamma^z \psi_R-\bar\psi_L\,\Gamma^z \psi_L
=\phi_\uparrow ^\dagger \phi_\uparrow-\phi_\downarrow^\dagger \phi_\downarrow
\label{axial}
\ea
Here we have defined the two-dimensional gamma matrices as $\Gamma^0=\sigma^1, \Gamma^z=-i \sigma^2, \Gamma^5=\sigma^3$.

Let us consider a generic chiral Lagrangian
\bea
\mc L_{4d}=\, \bar\Psi(i \gamma^\mu \partial_\mu)\Psi +\sum_{f=R,L}\mc L_{int,f}[\psi_f]
\label{4d_lag}
\ea
where the interaction term does not couple left and right sectors. Since $\gamma^0\gamma^\mu$ is also block-diagonal in the chiral basis, the left and right sectors are completely decoupled  and can be treated as independent ``flavors'' in the Lagrangian (\ref{4d_lag}). Each chiral sector has its own associated current $J_{R,L}=1/2(J^\mu\pm J^\mu_5)$ which are given by
\bea
J^0_f&=&\psi_f^\dagger\psi_f
\nn
J^1_f&=& \alpha_f \bar\psi_f\,\psi_f
\nonumber \\
J^2_f&=&i\bar\psi_f\,\Gamma^5\psi_f
\nonumber \\
J^3_f&=& \bar\psi_f\,\Gamma^z\psi_f
\label{left_right}
\ea
Here $f=L,R$, and $\alpha_R=-\alpha_L=1$. Even though the left and right currents seem to be completely independent, the conservation of vector current $\partial_\mu J^\mu=0$ still holds.

Now let us consider the lowest Landau level projection in the presence of a very strong magnetic field $B$ directed along $x_3$ direction as in Section \ref{subsec:1.1}. If the magnitude of the magnetic field is the largest scale in the problem, then the transition from the lowest Landau orbit to an excited orbit will be exponentially suppressed and motion along the transverse $x_1x_2$ plane will be frozen. The kinetic term in the Lagrangian (\ref{4d_lag}) then becomes:
 \bea
 \bar\Psi\gamma^\mu\partial_\mu\Psi&\rightarrow&\bar\Psi(\gamma^0\partial_0+\gamma^3\partial_3)\Psi=\sum_{f=R,L}\,\psi_f^\dagger\psi_f+\psi_f^\dagger\sigma^3\partial_3\psi_f=\sum_{f=R,L} \bar\psi_f\Gamma^M\partial_M\psi_f\nn
 \ea
by using the definition of two-dimensional gamma matrices above. This is the kinetic term of a two-dimensional Lagrangian.  Let us further assume that the dimensionally reduced interaction term $\mc L_{int,f}[\psi_f(z,t)]$ is invariant under the two-dimensional chiral rotation:
\bea
\psi_f(z,t)\rightarrow e^{i \Gamma^5 \zeta_f(z,t)}\psi_f(z,t)
\label{2d_chiral}
\ea
for an arbitrary function $\zeta_f(z,t)$. This two-dimensional notion of chirality generated by $\Gamma^5$ should not be confused with the four-dimensional chirality generated by $\gamma^5$. The former acts on each right and left subspinor separately. Also, the four-dimensional chiral Lagrangian (\ref{4d_lag}) does not have a term that couples right and left sectors so the system never develops a condensate such as $<\bar\psi_R\psi_L>$, and the four-dimensional chiral symmetry is never broken in our consideration. However, the dimensionally reduced system may exhibit dynamical breaking of the two-dimensional chiral symmetry (\ref{2d_chiral})\begin{footnote}{Spontaneous symmetry breaking in two dimensions is a delicate subject whose details are beyond the scope of our topic. It suffices to mention that in the limit of large number of flavors it can be realized, for example in the Gross-Neveu model \cite{Gross:1974jv}.}\end{footnote}. 
As an example let us consider the decomposition of the four-dimensional current-current interaction $\mc L_{int}=J^\mu J_\mu+J_5^\mu J_{5\mu}$.
The $(0, 3)$ components of the interaction become
\bea
J_f^0 J_{f 0}+J_f^3 J_{f 3}\rightarrow(\bar\psi_f\Gamma^M\psi_f)(\bar\psi_f\Gamma_M\psi_f)
\label{thirring}
\ea
In two-dimensions this corresponds to a Thirring interaction. Similarly, the transverse components
\bea
J_f^1 J_{f 1}+J_f^2 J_{f 2}\rightarrow(\bar\psi_f\psi_f)^2+(\bar\psi_f i\Gamma_5\psi_f)^2
\label{njl}
\ea
generate a chiral Gross-Neveu or Nambu-Jona-Lasinio (NJL) interaction in two-dimensions \cite{Gross:1974jv}. Note that both (\ref{thirring}) and (\ref{njl}) are invariant under (\ref{2d_chiral}) and exhibit chiral symmetry breaking in two-diemsions. The emergence of the Schwinger model from dimensional reduction of four-dimensional Maxwell-Chern-Simons theory investigated in Section \ref{sec:1.5} is another example. To sum up,  the four-dimensional currents (\ref{left_right}) are governed by a two-dimensional chiral Lagrangian
\bea
\mc L_{2d}=\sum_{f=R,L} \bar\psi_f\Gamma^M\del_M\psi_f+\mc L_{int,f}[\psi_f(x^M)]
\label{2d_lag}
\ea 
after dimensional reduction due to the strong magnetic field. 

We now show that the Lagrangian (\ref{2d_lag}) has certain model-independent properties concerning the  expectation values of the fermion bilinears $<\bar\psi_f\Gamma\psi_f>$. These are the building blocks of the expectation values of the four-dimensional currents (\ref{vector}, \ref{axial}). Note from (\ref{vector}) that $<J_f^0>$ and $<J^3_f>$ are expressed in terms of two-dimensional densities and currents of $\psi_f$, while the perpendicular components $<J_f^1>,<J_f^2>$ are expressed in terms of two-dimensional  scalar and pseudoscalar condensates of $\psi_f$.

\subsection{Life in two-dimensions}

Besides the magnetic field, the other necessary ingredient for chiral magnetic and/or separation effects is ambient charge density. In particular a nonzero baryon chemical potential $\mu$ leads to the chiral separation effect and a chiral (four-dimensional) chemical potential $\mu_5$ leads to the chiral magnetic effect. To keep the discussion general, let us consider nonzero chemical potential for both right and left sectors separately
\bea
\mu_R=\mu+\mu_5\neq0\quad,\quad\mu_L=\mu-\mu_5\neq0
\ea
which can be realized by adding the term $\sum_f \mu_f\psi_f^\dagger\psi_f$ to the Lagrangian. In two dimensions the same term can be generated by a special local chiral transformation with a {\it linear} dependence on the spatial coordinate $z$ in the exponent:
\bea
\psi_f^\prime=e^{-i \Gamma^5 \mu_f\, z}\psi_f.
\label{shift1}
\ea
To see this it is sufficient to observe
\bea
\bar\psi_f^\prime(i\,\Gamma^z\partial_z)\psi_f^\prime=\bar\psi_f(i\,\Gamma^z\partial_z)\psi_f+\mu_f\psi_f^\dagger\psi_f
\label{shift2}
\ea
where we used the gamma matrix identity $\Gamma^z\Gamma^5=\Gamma^0$. Since we assumed that the interaction term is invariant under any local chiral rotation of the form (\ref{2d_chiral}), the transformation (\ref{shift1}) does not generate any other term than the coupling to the chemical potential. Note that this feature  is special to two dimensions and does not generalize directly to higher dimensions. It is also crucial to have a chiral Lagrangian in four-dimensions. This would not hold for a massive fermion for instance. 

Let us start with the density $<\psi_f^\dagger\psi_f>$ and the reduced current along the $z$ direction $<\bar\psi_f \Gamma^z\psi_f>$ which constitute the four-dimensional density ($<J^0_f>$) and current along the magnetic field ($<J^3_f>$). In Section (\ref{subsec:1.3}) we have seen that the axial anomaly leads to an anomalous density in the presence of a chemical potential. In two-dimensional language, the axial anomaly
 \bea
\partial_M < j_{f5}^M> &=&\frac{e}{2 \pi}\epsilon^{MN}\partial_MA_{fN}\nn
<j_{f5}^M>&=& <\bar\psi_f \Gamma^M\Gamma^5\psi_f>=\epsilon^{MN} 
<\bar\psi_f \Gamma^N\psi_f>
 \ea
immediately reproduces $<\psi_f^\dagger\psi_f>=\mu_f/\pi$ once $e A_{f}^0$ is identified with the chemical potential $\mu_f$. Alternatively the same result can be obtained directly from path integral by observing the transformation (\ref{shift1}) applied to the renormalized charge density $<\bar\psi_f^{\prime\dagger}\psi_f^\prime>=0$ creates the anomalous term $\mu_f/\pi$ \cite{quarkyonic}.

Once the projection to the lowest Landau level is implemented, only the spin up component $\phi_\uparrow$ survives. Therefore for a right (left) handed spinor, only  the positive (negative) momentum component contributes to the anomalous density, reducing its value by a half:
\bea
<\psi_R^\dagger\psi_R>\quad \rightarrow\quad <R_+^*R_+>&=&\frac{e B}{2\pi}\frac{\mu_R}{2\pi}\qquad \text{(LLL projection)}\nn
<\psi_L^\dagger\psi_L>\quad \rightarrow \quad <L_-^*L_->&=&\frac{e B}{2\pi}\frac{\mu_L}{2\pi} \qquad\text{(LLL projection)}
\label{ano_density}
\ea
The overall factor $\frac{e B}{2\pi}$ is the density of the lowest Landau level in the transverse plane. There is no effect of the two-dimensional anomaly on the current $\bar\psi_f\Gamma^z\psi_f$ in general. However, the lowest Landau level projection leads to a nonzero anomalous contribution to the current as well:
\bea
&<\bar\psi_R\Gamma^z\psi_R>&=<\psi_R^\dagger\sigma_3\psi_R>\quad \rightarrow\quad<R_+^*R_+>=\frac{e B}{2\pi}\frac{\mu_R}{2\pi}\nn
&<\bar\psi_L\Gamma^z\psi_L>&=<\psi_L^\dagger\sigma_3\psi_L>\quad \rightarrow\quad-<L_-^*L_->=-\frac{e B}{2\pi}\frac{\mu_L}{2\pi}
\ea

 The strong magnetic field selects a particular spin. Provided that there is some excess charge in the medium, the magnetic field induces a current for each chiral sector along its direction. When we transform the right and left currents back to vector and axial currents, we see the chiral magnetic and chiral separation effects in their conventional form:
\bea
<J^3>=<J^0_5>=\frac{e B\,\mu_5}{2\pi^2}\,\,\quad\quad&\qquad\qquad& <J^3_5>=<J^0>=\frac{e B\,\mu}{2\pi^2}
\ea

Now consider another set of bilinears, those forming the scalar $<\bar\psi_f\psi_f>$ and pseudo-scalar $<\bar\psi_fi\Gamma^5\psi_f>$ condensates. They constitute the transverse components $<J^\perp_f>$ of the four-dimensional currents. In the semiclassical limit where the number of fermion flavors is large, the two-dimensional systems generically exhibit dynamical breaking of the chiral symmetry (\ref{2d_chiral}) and the system typically acquires a nonzero scalar condensate $<\bar\psi^\prime\psi^\prime>=m\neq0$ at zero chemical potential. Since this is also a mass term, it means there is a gap in the energy spectrum. The existence of the gap in the energy spectrum in one spatial dimension lowers the free energy at low temperatures as it ``pushes'' the Dirac sea further down in the energy spectrum. Once a finite chemical potential $\mu$ is turned on, all the states with energy lower than $\mu$ will be occupied and the optimal configuration would be to open a gap right around $\mu$ to push the occupied states down and lower the free energy. This is the celebrated Peierls instability \cite{peierls}.  and has broad consequences in condensed matter physics. Our assumption of the invariance of the interaction under the chiral transformation (\ref{shift1}) is sufficient to see that this scenario is indeed realized. The effect of (\ref{shift1}) on the associated Dirac Hamiltonian is
\bea
H \psi^\prime=-i \Gamma^5 \partial_z \psi^\prime+H_{int} \psi^\prime=(H-\mu)\psi. 
\ea
Therefore it is always possible to shift the energy spectrum, and  hence the gap, by $\mu$ with a chiral rotation. This is the relativistic version of the Peierls instability and it is proven for instance explicitly in the 2d Nambu-Jona-Lasinio model \cite{basar}, and is expected to be ubiquitous in two-dimensional systems with continuous chiral symmetry.

The chiral transformation (\ref{shift1}) mixes the scalar and pseudo-scalar condensates in the following way
\bea
\bar\psi_f^\prime\psi_f^\prime=\cos(2\mu_f z) \bar\psi_f\psi_f-\sin(2 \mu z) \bar\psi_f i\Gamma^5\psi_f
\label{cond_rot}
\ea 
so the existence of a nonzero condensate $<\bar\psi_f^\prime\psi_f^\prime>=m_f\neq0$ at zero chemical potential generalizes into finite chemical potential as
\bea
<\bar\psi\psi>= m\cos(2 \mu z)\quad,\quad<\bar\psi i\Gamma^5\psi>=-m \sin(2 \mu z)
\label{chiral_spiral}
\ea
since (\ref{cond_rot}) must hold for any $z$. This modulated scalar/pesudo-scalar condensate is referred to as the ``chiral spiral'' \cite{schon,Thies:2006ti}. In the 2d Nambu-Jona-Lasinio model the chiral spiral is indeed the thermodynamically preferred phase at low temperatures \cite{basar}. The chiral spiral translates into transverse components of the four-dimensional currents modulated in $z$:
\bea
<J^1_R>=c_R\,\cos(2\mu_R z+\phi_R)\,& \qquad \qquad & <J^1_L>=-c_L\,\cos(2\mu_L z+\phi_L)
\nn
<J^2_R>=c_R\,\sin(2\mu_R z+\phi_R)\,&\qquad \qquad & <J^2_L>=c_R\,\sin(2\mu_L z+\phi_L)
\ea
which we call the ``chiral magnetic spiral'' \cite{Basar:2010zd}. Here the amplitudes $c_{R,L}$ depend on the particular two-dimensional Lagrangian and are functions of the temperature, magnetic field and possibly other parameters in the model. However the chemical potential and space dependence of the currents is universal. $\phi_{R,L}$ are  relative phases of the left and right chiral condensates. 

It should be emphasized that, as opposed to the longitudinal currents, the chiral magnetic spiral mixes up and down spin components. This can be seen in the spin decomposition of the currents (\ref{vector}, \ref{axial}). In the lowest Landau projection, these pairings between spin up and down spinors correspond to excitations which can be described as a pairing of a particle with momentum $\mu_f$  and a hole with momentum $-\mu_f$. This is because the particle and hole have opposite charges and therefore opposite spins in the  lowest Landau level projection. Thus the excitation itself has momentum $\pm 2\mu_f$. This explains why the currents have the sinusoidal modulation in the $z$ direction. 

In heavy ion collisions, the chiral magnetic spiral can induce both out-of-plane and in-plane fluctuating charge asymmetries (the explicit separation of out-of-plane and in-plane fluctuations has been performed  \cite{Bzdak:2009fc} on the basis of STAR data \cite{:2009uh,:2009txa}). In the absence of topological fluctuations ($\mu_5 =0$), at finite baryon density ($\mu \neq 0$),  and in the chirally broken phase, the charge  current  has only transverse  components, and the charge asymmetry will fluctuate only in-plane.  It should be kept in mind that the presence of magnetic field increases the chiral transition temperature \cite{Gusynin:1995nb}. If topological fluctuations are present in the chirally broken phase (e.g. due to the presence of meta-stable $\eta'$ domains \cite{Kharzeev:1998kz}), the  CME current can be carried by the chiral magnetic spiral. The chiral magnetic spiral has also been seen in a holographic study \cite{holographiccms} in the framework of Sakai-Sugimoto model of holographic QCD \cite{Sakai:2004cn}.

%
%

\section{Fermions in an instanton and magnetic field background}

The previous discussion, along with many other papers, has presented the chiral magnetic effect as the flow of electrical charge as the result of some externally produced chirality imbalance, represented by a non-zero $\mu_5$. In this Section we consider the situation in which this chirality imbalance is produced not by an explicit $\mu_5$, or by a time-dependent $A_3$, but instead as a result of a topologically non-trivial gauge background like an instanton. Since quarks carry both electric and color charge, they couple to both electromagnetic and gluonic gauge fields. In this Section we discuss some features of the spectral problem for fermions in the combined background field of a strong magnetic field and an instanton \cite{Basar:2011by}.  To illustrate the effect most clearly we take a single instanton in $SU(2)$. We are motivated by situations in which quarks experience both types of fields, such as in dense astrophysical objects such as neutron stars and magnetars, and in heavy ion collisions such as those at RHIC and at CERN \cite{arXiv:0711.0950,arXiv:0907.1396,arXiv:1111.1949}.

 We are also motivated by recent lattice QCD analyses \cite{Buividovich:2009wi,Buividovich:2009my,Abramczyk:2009gb,tom,Braguta:2010ej,Tiburzi:2011vk}, which provide important numerical information about the Dirac spectrum in both QCD and magnetic field backgrounds. Analytically, while the effect of each individual background is very well known, their combined effect turns out to be quite intricate. In these lattice studies, certain matrix elements associated with chiral effects receive dominant contributions from zero-modes and near-zero-modes, so we pay particular attention to the low end of the spectrum, and show that certain generic features have a very simple analytic explanation.

\begin{figure}[htb]
\center\includegraphics[scale=0.2]{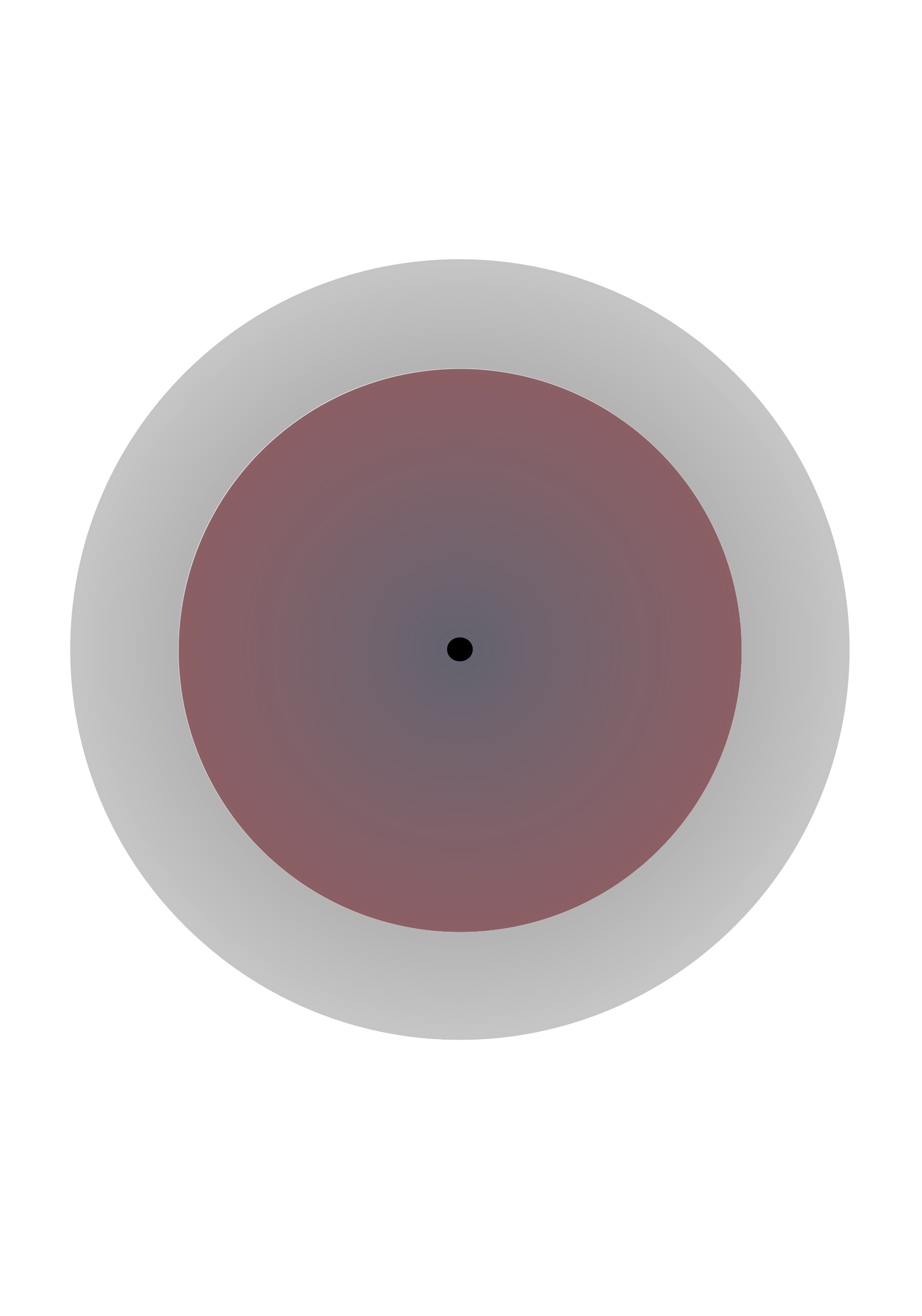}\qquad\qquad 
\includegraphics[scale=0.2]{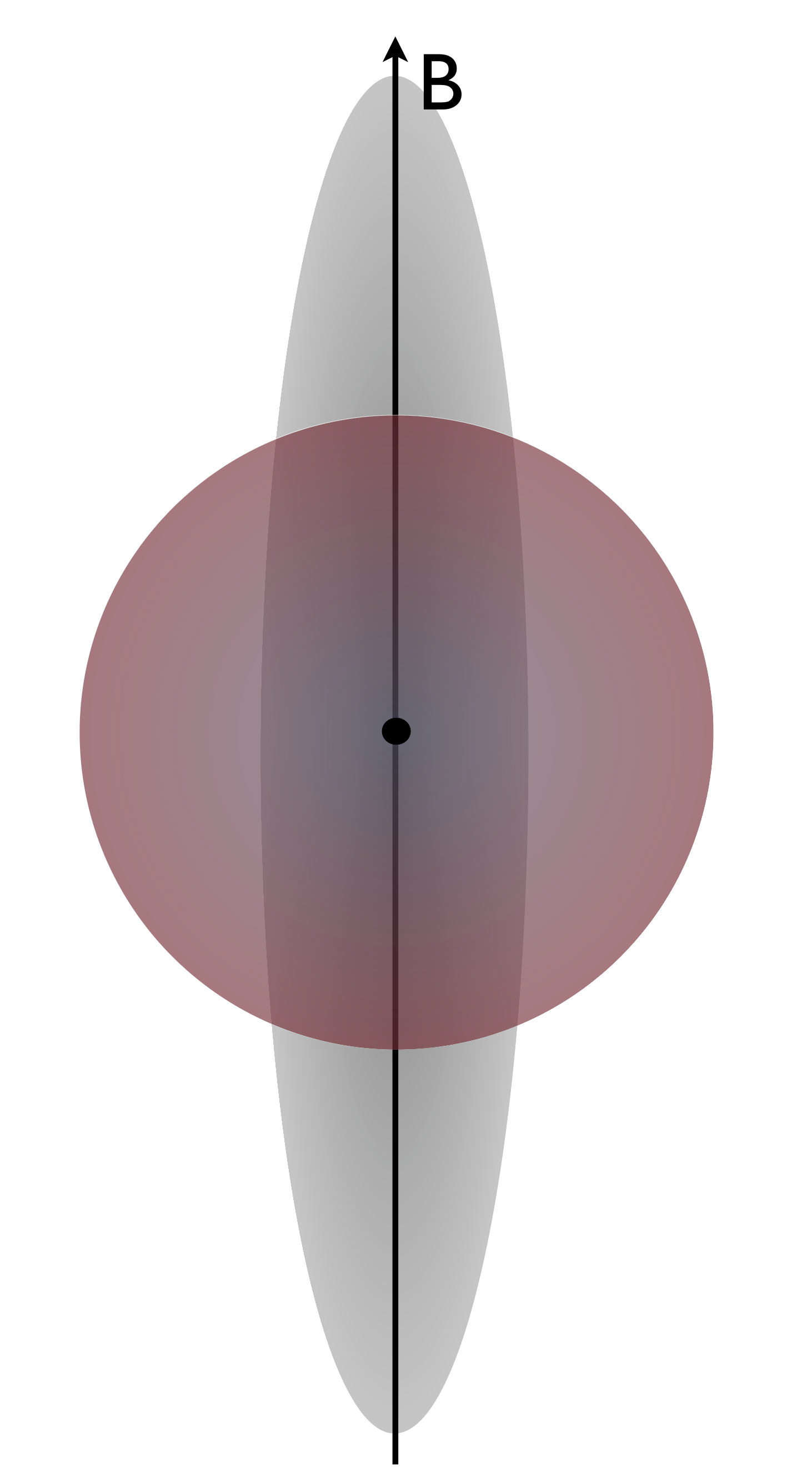}
\caption{A sketch of the topological charge density, $q\propto{\rm tr}\mc F_{\mu\nu}\tilde{\mc F}_{\mu\nu}$, for a single instanton [red], and the density of the quark zero mode [grey]. On the left, there is a single instanton, and both densities fall off as power laws, with $q$ falling off faster. On the right, with the introduction of a magnetic field, the topological charge density is unchanged but the zero mode density is distorted into an asymmetric shape, localized along the direction of the strong magnetic field.
}
\label{fig4}
\end{figure}

As discussed already, a magnetic field introduces a Landau level structure to the fermion spectrum, in which the zero modes  of the associated two-dimensional Euclidean Dirac operator have definite spin, aligned along the magnetic field  \cite{aharonov-casher,novikov}. For a constant magnetic field on a torus, as appropriate for  lattice QCD analysis, this has been studied recently in \cite{Giusti:2001ta,Tenjinbayashi:2005sy,AlHashimi:2008hr}. The appropriate formalism is that of the magnetic translation group \cite{Zak:1964zz}. In a gluonic field with nontrivial topological charge (for example, an instanton), the fermion spectrum of the four-dimensional Euclidean Dirac operator also has zero modes, with chiralities determined locally by the local topological charge of the gauge field \cite{'tHooft:1976fv,Schwarz:1977az,Kiskis:1977vh,Jackiw:1977pu,rubakov}. For a single instanton the fermion spectral problem has a conformal symmetry \cite{Jackiw:1976dw,Chadha:1977mh}, and the zero modes are localized on the instanton, falling off as a power law with Euclidean distance. The conformal symmetry is broken by the introduction of a magnetic field, and now the zero modes develop an asymmetry, falling off in Gaussian form in the plane transverse to the B field, but as a power law in the other two directions. This basic asymmetry is an important feature of the phenomena of magnetic catalysis \cite{Gusynin:1995nb} and the chiral magnetic effect \cite{Kharzeev:2004ey,arXiv:0706.1026,arXiv:0711.0950,Fukushima:2008xe,Kharzeev:2009fn}, as sketched in Fig. \ref{fig4}.

\subsection{Euclidean Dirac Operator}

To discuss a combined background of an instanton and a static magnetic field we use Euclidean Dirac matrices, instead of the Minkowski ones used in the previous Sections. Our conventions follow those of \cite{Jackiw:1977pu}, expressing the $4\times 4$ Dirac matrices,
$\gamma_\mu$, for $\mu=1, 2, 3, 4$, in terms of the $2\times 2$ matrices $\alpha_\mu=(\mathbb 1, -i\vec{\sigma})$ and $\bar\alpha_\mu=(\mathbb 1, i\vec{\sigma})=\alpha_\mu^\dagger$, [here $\vec{\sigma}$ are the usual $2\times 2$ Pauli matrices]:
\bea
\gamma_\mu=
\begin{pmatrix}
0&\alpha_\mu\\
\bar\alpha_\mu & 0
\end{pmatrix}
\qquad, \qquad 
\gamma_5=
\begin{pmatrix}
\mathbb 1&0\\
0& -\mathbb 1
\end{pmatrix}
\label{gammas}
\ea
Thus, the Euclidean Dirac operator can be expressed as
\bea
\slashed{D}= \left(\begin{matrix} 0 & \alpha_\mu D_\mu \\\bar\alpha_\mu D_\mu & 0  \end{matrix} \right) \equiv \left(\begin{matrix} 0 & D\\-D^\dagger & 0  \end{matrix} \right) 
\label{dirac}
\ea
where the covariant derivative, $D_\mu=\partial_\mu-i \mc A_\mu$, is written with a hermitean gauge field,  $\mc A_\mu$, and $x_4$ is the Euclidean time coordinate. We write the gauge field $\mc A_\mu$ as a sum of a non-abelian part, $A_\mu$, and an abelian part, $a_\mu$:
\bea
\mc A_\mu=A_\mu+a_\mu
\label{gauge}
\ea
with the respective coupling constants absorbed into the gauge fields.
The Dirac operator is anti-hermitean, so we write (with $\lambda$ real)
\bea
i\slashed{D}\,\psi_\lambda=\lambda \,\psi_\lambda
\label{lambda}
\ea
Since $\{\gamma_5, \slashed D\}=0$, we can take $\lambda$ in (\ref{lambda}) to be non-negative, with the negative eigenvalue solutions simply given by $\psi_{-\lambda}=\gamma_5\psi_\lambda$. This means that we can effectively discuss the zero modes ($\lambda=0$) separately, and for the nonzero modes  ($\lambda\neq0$) we consider the squared operator:
\bea
\left(i\slashed{ D}\right)^2\psi_\lambda
=
\begin{pmatrix}
DD^\dagger &0\\
0& D^\dagger D
\end{pmatrix}
\psi_\lambda=\lambda^2\psi_\lambda
\label{eig}
\ea
The positive chirality sector, $\chi=+1$, is described by the operator $DD^\dagger$, while the negative chirality sector, $\chi=-1$, is described by the operator $D^\dagger D$. We can write these operators as
\bea
\chi=+1 : \qquad DD^\dagger &=&-D_\mu^2-\frac{1}{2}\mc F_{\mu\nu}\bar\sigma_{\mu\nu}
\label{dd-1}\\
\chi=-1 : \qquad D^\dagger D &=&-D_\mu^2-\frac{1}{2}\mc F_{\mu\nu}\sigma_{\mu\nu}
\label{dd-2}
\ea
We have used $[D_\mu, D_\nu]=-i\mc F_{\mu\nu}$, where $\mc F_{\mu\nu}$ is the field strength associated with the gauge field $\mc A_\mu$, and the spin matrices $\bar\sigma_{\mu\nu}$ and $\sigma_{\mu\nu}$ are defined as
\bea
\bar\sigma_{\mu\nu}=\frac{1}{2i}\left(\alpha_\mu\bar\alpha_\nu-\alpha_\nu\bar\alpha_\mu\right)\quad,\quad\sigma_{\mu\nu}=\frac{1}{2i}\left(\bar\alpha_\mu\alpha_\nu-\bar\alpha_\nu\alpha_\mu\right)
\label{spin}
\ea
In (\ref{dd-1},\ref{dd-2}) we have used the properties \cite{Jackiw:1977pu}: $\bar\alpha_\mu\alpha_\nu=\delta_{\mu\nu}+i \sigma_{\mu\nu}$, and $\alpha_\mu\bar\alpha_\nu=\delta_{\mu\nu}+ i \bar\sigma_{\mu\nu}$.

For non-zero modes [i.e., solutions to (\ref{eig}) with $\lambda\neq 0$], the operators $DD^\dagger$ and $D^\dagger D$ have identical spectra, for any background field. This is simply because we have an invertible map: suppose the 2-component spinor $v$ satisfies $D^\dagger D v=\lambda^2 v$. Then $u= D v$ is clearly an eigenfunction of the other operator, $D  D^\dagger$, with precisely the same eigenvalue: $D  D^\dagger u=D D^\dagger D v =\lambda^2 u$. Similarly, if $u$ satisfies $D D^\dagger u=\lambda^2 u$, then $v=D^\dagger u$ is an eigenstate of $D^\dagger D$ with the same eigenvalue. 
Thus, when $\lambda\neq 0$, we can write the 4-component spinor solution in the form
\bea
\psi_\lambda=
\begin{pmatrix}
u_\lambda \\
-\frac{i}{\lambda}\,D^\dagger u_\lambda
\end{pmatrix}
\qquad {\rm where} \qquad DD^\dagger u_{\lambda}=\lambda^2 u_\lambda
\label{spinor1}
\ea
or in the form
\bea
\psi_\lambda=
\begin{pmatrix}
\frac{i}{\lambda}D v_\lambda\\
v_\lambda
\end{pmatrix}
\qquad {\rm where} \qquad D^\dagger D v_\lambda=\lambda^2 v_\lambda
\label{spinor2}
\ea
This is true for any background field: non-abelian, abelian, or both. 

\subsection{Magnetic field background}\label{magback}

For a constant (abelian) magnetic field, of strength $B$,  pointing in the $x_3$ direction, we have an abelian field strength $f_{12}=B$, and so we find 
\bea
D^\dagger D=DD^\dagger =-D_\mu^2-B \sigma_3
\label{dmag}
\ea
where we have used the fact that $\bar\sigma_{12}=\sigma_{12}=\sigma_3$. Due to the subtraction term, $-B\sigma_3$, it is possible to have zero modes, and since $D D^\dagger=D^\dagger D$ these zero modes occur in each chiral sector. More explicitly, we can make a Bogomolnyi-style factorization similar to (\ref{bogom}) and write 
\bea
-D_\mu^2-B \sigma_3
&=&-\partial_3^2-\partial_4^2-\left(D_1\mp i D_2\right)\left(D_1\pm i D_2\right)\pm B -B\sigma_3
\nn
&=&-\partial_3^2-\partial_4^2- D_\mp D_\pm \pm B-B\,\sigma_3
\label{bog}
\ea
For zero modes, we take $\partial_3=\partial_4=0$, and with $B>0$ we choose the upper signs to ensure normalizable modes. For example, in the symmetric gauge where the abelian gauge field
\bea
a_\mu=\frac{B}{2}(-x_2, x_1, 0, 0)
\label{symmetric}
\ea
 the zero modes can be expressed  in terms of the normalizable solutions to $\left(D_1+ iD_2\right)u=0$:
\bea
\psi_0=g(z_1)e^{-B|z_1|^2/4}
\begin{pmatrix}
1\\0\\0\\0
\end{pmatrix}
\qquad {\rm or} \qquad 
\psi_0=g(z_1)e^{-B|z_1|^2/4}
\begin{pmatrix}
0\\0\\1\\0
\end{pmatrix}
\ea
Here $g(z_1)$ is a holomorphic function of the complex variable $z_1=(x_1+i x_2)$. Both sets of zero modes have spin up, aligned along the $B$ field; this is just the familiar lowest Landau level projection onto spin up states. Note also that the zero modes have the characteristic Gaussian factor in the $(x_1, x_2)$ plane, transverse to the direction of the magnetic field. This factor is the origin of the distortion sketched in the right frame of Fig. \ref{fig4}.

The number of zero modes per unit two-dimensional area [in the $(x_1,x_2)$ plane] is given by the Landau degeneracy factor, the magnetic flux per unit area: $B/(2\pi)$. In fact, even for an inhomogeneous magnetic field $B(x_1, x_2)$, pointing in the $x_3$ direction, the number of zero modes [of each chirality] is determined by the integer part of the magnetic flux (this is the essence of the Aharonov-Casher theorem \cite{aharonov-casher}). For example, on a torus \cite{novikov}:
\bea
N_+=N_-=\frac{1}{2\pi}\int d^2 x\,  B
\label{aharonov}
\ea
The higher Landau level states are the same for both spins, because  $(-D_-D_+ +B)$ and $(-D_+D_--B)$ have identical spectra, apart from the lowest level, which only has spin aligned along the magnetic field. 

\subsection{Instanton background}

For an instanton field, $A_\mu$, the (non-abelian) field strength $F_{\mu\nu}$ is self-dual [that is: $F_{\mu\nu}=\tilde{F}_{\mu\nu}$, where the dual tensor is defined: $\tilde{F}_{\mu\nu}\equiv\frac{1}{2}\epsilon_{\mu\nu\alpha\beta}{F}_{\alpha\beta}$]. Then the anti-self-duality property of $\bar\sigma_{\mu\nu}$ [that is: $\bar\sigma_{\mu\nu}=-\frac{1}{2}\epsilon_{\mu\nu\rho\sigma}\bar\sigma_{\rho\sigma}$] implies:
\bea
\chi=+1 : \qquad DD^\dagger &=&-D_\mu^2\\
\chi=-1 : \qquad D^\dagger D &=&- D_\mu^2-\frac{1}{2}\,F_{\mu\nu}\sigma_{\mu\nu}
\label{ddd-inst}
\ea
Since $-D_\mu^2$ is a positive operator, this means that for an instanton background there can be no zero mode in the positive chirality sector. On the other hand, due to the subtraction term, $-F_{\mu\nu}\sigma_{\mu\nu}$, in $D^\dagger D$, it is possible to have a zero eigenvalue solution in the negative chirality sector, and it has the form
\bea
\psi_0=\begin{pmatrix}
0\\v
\end{pmatrix} \qquad , \quad {\rm where} \qquad Dv=0
\label{zeromode}
\ea
[For an anti-instanton, an anti-self-dual field with $F_{\mu\nu}=-\tilde{F}_{\mu\nu}$,  the zero mode lies in the positive chirality sector, because $\sigma_{\mu\nu}$ is self-dual: $\sigma_{\mu\nu}=\frac{1}{2}\epsilon_{\mu\nu\rho\sigma}\sigma_{\rho\sigma}$.] For a general non-abelian gauge field $A_\mu$, which is neither self-dual nor anti-self-dual, the Atiyah-Singer  index theorem \cite{aps,rubakov} states that the difference between the number of positive and negative chirality zero modes is given by the topological charge of the gauge field:
\bea
N_+-N_-=- \frac{1}{32\pi^2}\int d^4x\,
F^a_{\mu\nu}\tilde{F}^a_{\mu\nu}
\label{atiyah}
\ea
Here we have written $F_{\mu\nu}=F_{\mu\nu}^a T^a$, with generators normalized as 
${\rm tr}(T^a T^b)=\frac{1}{2}\delta^{ab}$.
For gauge group $SU(2)$, with fermions in the defining representation, we take generators $T^a=\frac{1}{2}\tau^a$ in terms of the Pauli matrices $\vec{\tau}$, and we can  write the single instanton gauge field \cite{Belavin:1975fg}, centered at the origin,  in the regular gauge as 
\bea
A_\mu^a=2\frac{ \eta^a_{\mu\nu}\, x_\nu}{x^2+\rho^2}
\ea
where $\rho$ is the instanton scale parameter, and $\eta^a_{\mu\nu}$ is the self-dual  't Hooft tensor \cite{'tHooft:1976fv,Jackiw:1977pu}. The topological charge density is 
\bea
q(x)=\frac{1}{32\pi^2}\, F^a_{\mu\nu}\tilde{F}^a_{\mu\nu}=\frac{192\, \rho^4}{(x^2+\rho^2)^4}
\label{q}
\ea
There is a single zero mode \cite{'tHooft:1976fv,Schwarz:1977az,Kiskis:1977vh,Brown:1977bj,rubakov}, also localized at the origin, with density:
\bea
|\psi_0|^2=
\frac{64\, \rho^2}{(x^2+\rho^2)^3}
\label{izm}
\ea
These densities both fall off as power laws, with scale set by $\rho$, but the topological charge density is more localized, as indicated in the left-hand frame of Fig. \ref{fig4}. The nonzero modes are given by (\ref{spinor1}) or (\ref{spinor2}), and we note that the spectra are identical in each chiral sector, apart from the zero modes.

\subsection{Combined instanton and magnetic field background}

Physically,  an instanton field projects the zero modes onto a definite chirality, while a constant magnetic field projects the zero modes onto definite spin, aligned along the direction of the magnetic field. When we combine the two background fields, both a non-abelian instanton field $F_{\mu\nu}$ and an abelian magnetic field $f_{12}=B$, there is a competition between the two projection mechanisms, and the outcome depends on their relative magnitude, as we show below. Technically speaking, the instanton zero mode has a specific ansatz form that unifies space-time and color indices, while the magnetic zero modes have a natural holomorphic structure, and these two different ansatz forms do not match one another. The competition between these two ansatz forms makes the combined problem nontrivial. For an instanton field, since the field falls off as a power law, all eigenmodes also fall off with power law behavior. On the other hand, once a constant magnetic field is introduced, for example in the gauge (\ref{symmetric}), all the eigenstates (even those in the higher Landau levels) have a Gaussian factor
$\exp(-B|z_1|^2/4)$ that localizes the modes near the axis of the magnetic field. This is the reason for the distorted density in the right-hand frame of Fig. 1. In the extreme strong magnetic field limit this leads to a dimensional reduction to motion along the magnetic field, with interesting physical consequences such as magnetic catalysis \cite{Gusynin:1995nb} and the chiral magnetic effect \cite{Kharzeev:2004ey,Fukushima:2008xe,Kharzeev:2009fn}.

Concerning zero modes, we begin with a simple but important comment: in the index theorem (\ref{atiyah}), the magnetic field makes no contribution, since with the field strength decomposed into its non-abelian and abelian parts, $\mc F_{\mu\nu}=F_{\mu\nu}+f_{\mu\nu}$, we have
\bea
{\rm tr}\left(\mc F_{\mu\nu}\tilde{\mc F}_{\mu\nu}\right)&=&{\rm tr}\left(F_{\mu\nu}\tilde{F}_{\mu\nu}\right)+({\rm dim})\,f_{\mu\nu}\tilde{f}_{\mu\nu}\\
&=&{\rm tr}\left(F_{\mu\nu}\tilde{F}_{\mu\nu}\right)
\ea
where dim is the dimension of the Lie algebra representation of the non-abelian gauge fields. The cross terms vanish since the Lie algebra generators $T^a$ are traceless, and the $f_{\mu\nu}\tilde{f}_{\mu\nu}$ term vanishes since there is no abelian electric field. For example, if there is no nonabelian field, just an abelian magnetic field, then the topological charge clearly vanishes, and  the index theorem (\ref{atiyah})  is consistent with the fact that $DD^\dagger =D^\dagger D$ for an abelian magnetic background (recall (\ref{dmag})), so that there are the same number of zero modes in each chiral sector.
Now, with both background fields present, we find
\bea
DD^\dagger &=&-D_\mu^2-B\sigma_3\\
D^\dagger D &=&-D_\mu^2-\frac{1}{2}\,F_{\mu\nu}\sigma_{\mu\nu}-B\sigma_3
\label{ddd-both}
\ea
Notice that the eigenvalues of $DD^\dagger$ are simply those of the scalar operator $-{D}_\mu^2$, with a spin term $\pm B$, as can be seen clearly in Figure \ref{fig5}.
The fact that there is a subtraction term from the positive operator $-D_\mu^2$ in both chirality sectors tells us that it is possible to have zero modes for each chirality, but their number will depend on the relative magnitude of $F$ and $B$. In the next Section  we study a specific model where we can quantify this precisely. Another important implication is that we may also have some "near-zero-modes", where the $F$ and $B$ subtractions do not exactly cancel the lowest eigenvalue of $-D_\mu^2$, but lower the eigenvalue of $DD^\dagger$ or $D^\dagger D$ to near zero.

\subsection{Large instanton limit: Covariantly constant $SU(2)$ instanton and constant abelian magnetic field}

In the very strong magnetic field limit, where the magnetic length, $1/\sqrt{B}$, is small compared to the instanton size $\rho$, we expect  a significant distortion of instanton modes and currents. In this limit we can make a simple approximation that reduces the problem to a completely soluble system.

In the large instanton limit, we expand the instanton gauge field as:
\bea
A_\mu^{a}\approx \frac{2}{\rho^2 } \eta^a_{\mu\nu} x_\nu+\dots
\label{leading}
\ea
To leading order in such a  derivative expansion, the non-abelian gauge configuration $A_\mu^a(x)$ is self-dual and has covariantly constant field strength: $F_{\mu\nu}^a=-\frac{4}{\rho^2} \,\eta_{\mu\nu}^a$. 
In this limit we can make an $SU(2)$ "color" rotation, along with a choice of Lorentz frame,  to make the instanton field diagonal  in the color space (we choose the $\tau^3$ direction), so that the field is self-dual, covariantly constant and quasi-abelian. 
 Defining the instanton scale $F=\frac{2}{\rho^2}$, 
 the combined gauge field, including also the abelian magnetic field as in (\ref{gauge}), can be written as: 
\bea
\mc A_\mu=-\frac{F}{2}(-x_2, x_1, -x_4, x_3)\tau^3+\frac{B}{2}(-x_2, x_1, 0, 0)\mathbb{1}_{2\times2}
\label{gauge2}
\ea
This gauge field is fully diagonal and moreover is linear in $x_\mu$, so the problem is analytically soluble (this is the basic premise of the derivative expansion).
The only nonzero entries of the field strength tensor are
\bea
\mc F_{12}&=&-F\tau^3+B\mathbb{1}=
\begin{pmatrix}
B-F &0\cr
0&B+F
\end{pmatrix}
\nonumber\\
\mc F_{34}&=&-F\tau^3
=
\begin{pmatrix}
-F &0\cr
0&+F
\end{pmatrix}
\ea
In the absence of the magnetic field the field strength is self-dual, $\mc F_{12}=\mc F_{34}$, but a nonzero magnetic field breaks this symmetry. The topological charge density is (recall the normalization of the generators)
\bea
\frac{1}{32\pi^2} \,\mc F_{\mu\nu}^a \tilde{\mc F}_{\mu\nu}^a=\frac{4(2F)^2}{32\pi^2}=\frac{F^2}{2\pi^2}
\label{tc}
\ea

A natural question to ask is: in such a constant field strength background, the wave functions have a {\it Gaussian} spatial dependence in the plane transverse to the direction of the field, characteristic of the Landau problem, so how can we recover the power-law dependence of the zero modes in an instanton background? This happens as follows. Recall \cite{rubakov} that with the appropriate ansatz for the zero mode, the zero mode equation reduces to a first-order radial equation
\begin{eqnarray}
\psi^\prime_0=-\frac{3r}{1+r^2}\, \psi_0 \quad \Rightarrow \quad \psi_0=\frac{1}{(1+r^2)^{3/2}}
\label{reduced}
\end{eqnarray}
where $r$ is the Euclidean distance. At short and long distances, this zero mode behaves as
\begin{eqnarray}
\psi_0 &\sim& 1-\frac{3}{2}r^2+\dots \qquad\qquad , \qquad r\to 0\\
\psi_0 &\sim& \frac{1}{r^3}+\dots \qquad\qquad\qquad  , \qquad r\to \infty
\label{asymp}
\end{eqnarray}
On the other hand, if we make the above choice (\ref{gauge2}) of a constant field to represent a large instanton, we have instead the zero mode equation
\begin{eqnarray}
\psi^\prime_0=- 3r\, \psi_0 \quad \Rightarrow \quad \psi_0=e^{-3 r^2/2}
\label{constant}
\end{eqnarray}
which has the correct short-distance behavior but which is Gaussian rather than power law at large distances. 
\begin{figure}[htb]
\includegraphics[scale=1]{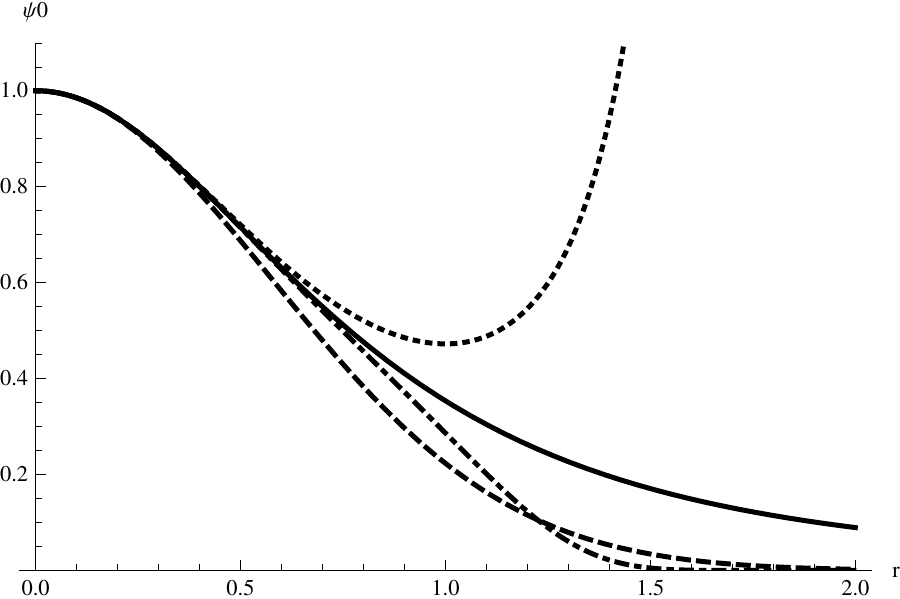}
\label{fig5}
\caption{Succesive approximations in the derivative expansion of the instanton background interpolate between the exponential form of the zero mode [dashed line] to the power-law form [solid line] of the exact result.}
\end{figure}

We can recover the correct power-law behavior by restoring the instanton size parameter, including the sub-leading terms in (\ref{leading}),  and expanding the zero mode equation as
\begin{eqnarray}
\psi^\prime_0=-3r\left(1-r^2+r^4-r^6+r^8-\dots\right)\, \psi_0 
\label{reduced2}
\end{eqnarray}
The leading term represents the leading term of the derivative expansion of the instanton field, and is  solved by the Gaussian factor. Absorbing this Gaussian factor by writing $\psi_0=e^{-3 r^2/2}\chi_0(r)$, the resulting equation for $\chi_0(r)$ is
\begin{eqnarray}
\chi^\prime_0=3r\left(r^2-r^4+r^6-r^8+\dots\right)\, \chi_0 
\label{reduced3s}
\end{eqnarray}
which suggests writing $\chi_0=e^{3r^4/4} \phi_0(r)$. Continuing this process order by order in the derivative expansion, we obtain 
\begin{eqnarray}
\psi_0(r)=\exp\left[-\frac{3}{2}r^2+\frac{3}{4}r^4-\frac{3}{6}r^6+\frac{3}{8}r^8-\dots\right]
=\exp\left[-\frac{3}{2}\ln(1+r^2)\right]
\end{eqnarray}
thereby recovering the correct power-law decay from the derivative expansion of the instanton background, as illustrated in Figure \ref{fig5}.

\subsection{Dirac spectrum in the strong magnetic field limit}

To study the Dirac spectrum with both a magnetic field and an instanton  we consider the $2\times 2$ operators $DD^\dagger$ and $D^\dagger D$  in (\ref{dd-1},\ref{dd-2}). Notice first that
\bea
\mc F_{\mu\nu}\,\bar\sigma_{\mu\nu}&=&2\left(\mc F_{12}-\mc F_{34}\right)\sigma_3
\\
\mc F_{\mu\nu}\,\sigma_{\mu\nu}&=&2\left(\mc F_{12}+\mc F_{34}\right)\sigma_3
\ea
It is convenient to factor the 4-dimensional Euclidean space and consider separately the $(x_1, x_2)$ plane and the $(x_3, x_4)$  plane.
Then in the $(x_1, x_2)$ plane we have a (relativistic) Landau level problem with effective field strength $(B-F)$ in the $\tau^3=+1$ sector, and with effective field strength $(B+F)$ in the $\tau^3=-1$ sector. In the $(x_3, x_4)$ plane we also have a (relativistic) Landau level problem, now with effective field strength $-F$ in the $\tau^3=+1$ sector, and with effective field strength $F$ in the $\tau^3=-1$ sector. In the $(x_1, x_2)$ plane the sign of the effective field strength depends on which of $B$ or $F$ is larger, and  in the strong $B$ field limit, both $B\pm F$ are positive.

When $B>F$, both $(B-F)$ and $(B+F)$ are positive. Thus, each color component of $\mc F_{12}$ is associated with a positive "magnetic" field. On the other hand, for $\mc F_{34}$, the $\tau^3=+1$ sector has  a negative field strength, while the  $\tau^3=-1$ sector has  a  positive  field strength. 

We first consider the $\tau^3=+1$ case. Then $\mc F_{12}=(B-F)$, $\mc F_{34}=-F$,
$\mc F_{\mu\nu}\bar\sigma_{\mu\nu}=2B\sigma_3$, and $\mc F_{\mu\nu}\sigma_{\mu\nu}=2(B-2F)\sigma_3$. With  a positive field strength the normalizable zero state is given by $\left(D_1+i D_2\right)u=0$.
But since  $\mc F_{34}$ is negative, we factorize the corresponding covariant derivatives in the opposite order, in order to obtain a normalizable state annihilated by $\left(D_3-iD_4\right)$. Thus, we have, for chirality $\chi=\pm 1$, respectively:
\bea
D D^\dagger
&=& -\left(D_1-iD_2\right)\left(D_1+iD_2\right) 
- \left(D_3+iD_4\right)\left(D_3-iD_4\right)
+B-B \sigma_3\nn
D^\dagger D
&=& -\left( D_1-iD_2\right)\left(D_1+iD_2\right) 
- \left(D_3+iD_4\right)\left(D_3-iD_4\right)
+B-(B-2F) \sigma_3\nn
\label{bf+-}
\ea
This shows that there is a zero mode,  when the spin term $B \sigma_3$ cancels the $B$ term from the Bogomolnyi factorization of the covariant derivative term. This occurs  in the positive chirality sector, $\chi=+1$, and  with spin up: $\sigma_3=+1$.

Now consider the $\tau^3=-1$ case. 
Then $\mc F_{12}=(B+F)$, $\mc F_{34}=F$,
$\mc F_{\mu\nu}\bar\sigma_{\mu\nu}=2B\sigma_3$, and $\mc F_{\mu\nu}\sigma_{\mu\nu}=2(B+2F)\sigma_3$.
All field strengths are positive, so we  write, for chirality $\chi=\pm 1$, respectively:
\bea
D D^\dagger
&=& -\left(D_1-iD_2\right)\left(D_1+i\mc D_2\right) 
- \left(D_3-iD_4\right)\left(D_3+iD_4\right)
+(B+2F)-B \sigma_3
\nn
D^\dagger D
&=& -\left(D_1-iD_2\right)\left(D_1+iD_2\right) 
- \left(D_3-iD_4\right)\left(D_3+iD_4\right)
+(B+2F)-(B+2F) \sigma_3
\nn
\label{bf--}
\ea
This shows that there is a zero mode, but now in the opposite chirality sector, $\chi=-1$, and also with spin up: $\sigma_3=+1$.

To summarize: when $B>F$, the $\tau_3=+1$ color sector has  spin up zero modes with positive chirality, while the $\tau_3=-1$ color sector has  spin up zero modes with negative chirality. 
We can count the number of zero modes in each chirality sector by simply taking the product of the Landau degeneracy factors for the $(x_1, x_2)$ and $(x_3, x_4)$ planes, with the corresponding effective magnetic field strengths. Therefore, 
the corresponding Landau degeneracy factors give the zero-mode number densities (i.e, the number per unit volume):
\bea
\chi=+1:\qquad n_+&=&\frac{(B-F)}{2\pi}\,\frac{F}{2\pi} \qquad (\tau_3=+1\quad, \quad \sigma_3=+1)
\label{bf-landau-1}\\
\chi=-1:\qquad n_-&=&\frac{(B+F)}{2\pi}\,\frac{F}{2\pi}  \qquad (\tau_3=-1\quad, \quad \sigma_3=+1)
\label{bf-landau-2}
\ea
The index (density) is the difference, 
\bea
n_+-n_-=-\frac{F^2}{2\pi^2}
\ea
in agreement with the general index theorem (\ref{atiyah}), in view of (\ref{tc}). 
We also note that the total number density of zero modes
\bea
n_++n_-=\frac{BF}{2\pi^2}
\ea
is linearly proportional to the magnetic field strength $B$. This is in agreement with numerical lattice gauge theory results \cite{tom}.

It is worth emphasizing that if torus boundary conditions $x_\mu \sim x_\mu+L_\mu$ are imposed, then the fluxes are quantized as $BL^2=2\pi M$ and $FL^2=2\pi N$. Here $M$ and $N$ are positive integers. Consequently the index and total number of zero modes are given by also by integers
\bea
&{\rm index}(\slashed{ D})\equiv N_+-N_- =(N-M)M-(N+M)M=-2M^2&\nn
&\text{total number of zero modes}=(N+M)M+(N-M)=2NM&
\label{torus_zero}
\ea
Moreover, the instanton solution with constant field strength (\ref{leading}) is shown to be an exact solution of Yang-Mills equations on a four torus \cite{'tHooft:1981sz,vanBaal}. Even with the inclusion of the magnetic field, (\ref{gauge2}) stays as a solution. Therefore on a on four-torus, there is no restriction on the magnitude of $B$ and $F$ (i.e. we do not have to consider a large instanton) and our counting of the zero modes (\ref{torus_zero}) becomes exact for any integer $M$ and $N$.

\subsection{Physical picture: competition between spin and chirality projection}

These results lead to the following simple physical picture. The instanton tries to generate a chirality imbalance but is neutral to the spin, whereas the magnetic field tries to generate a spin imbalance but does not affect the chirality. Depending on which is stronger, the zero modes have either a definite spin with a chirality imbalance ($B>F$), or a definite chirality with a spin imbalance ($F>B$). 
Also we see that in the former case, the {\it total} number of zero modes scales with $B$ and is not equal to the index.

More explicitly, for the $B>F$ case, consider starting with just a strong magnetic field $B$, later  turning on a weak instanton field.  Without the instanton field, the zero modes and their degeneracy are given by the Aharonov-Casher theorem (\ref{aharonov}), so that the zero mode density is the Landau degeneracy factor $B/(2\pi)$ for each chirality sector. 
All the zero modes are spin up, as is familiar for the lowest Landau level (see Fig. \ref{fig2}). 
There is an equal number of positive and negative chirality zero-modes, 
 which is consistent with the index theorem, since the topological charge vanishes for a constant $B$ field. 
Now consider turning on an instanton field $F$, with $B>F>0$. We see from (\ref{bf-landau-1},\ref{bf-landau-2}) that the effect of the instanton is to flip some of the chiralities: $\left(\frac{F}{2\pi}\right)^2$ positive chirality modes become negative chirality modes,  leading to a chirality imbalance of $\frac{F^2}{2\pi^2}$, in agreement with the index theorem (\ref{atiyah}). On the other hand, the total number of zero modes, $\frac{BF}{2\pi^2}$, grows linearly with the magnetic field when $F$ is nonzero.

\subsection{Matrix Elements and Dipole Moments}

The computation of matrix elements is significantly simplified by using the Euclidean version of the Schur decomposition (\ref{s-prop1}, \ref{s-prop2}) used earlier to relate the chiral magnetic effect to the two- and four-dimensional axial anomalies. Thus,  introducing a small quark mass $m$,  the propagator of the
 Dirac operator $\slashed{\mc D}+m$ is given by its Schur decomposition:
 \bea
 \frac{1}{\slashed{D}+m}=
 \begin{pmatrix}
 \frac{m}{m^2+DD^\dagger} &  \frac{-1}{m^2+DD^\dagger} \, D\cr
 \frac{1}{m^2+D^\dagger D} \, D^\dagger &  \frac{m}{m^2+D^\dagger D}
 \end{pmatrix}
 \eea
 Note that $DD^\dagger$ and $D^\dagger D$ have identical spectra, except for possible zero modes, so they can be viewed as square operators (matrices) of different dimension, as is clear when they are diagonalized in their respective eigenspaces. For a simple algebraic illustration, take $D$ to be the 2-component column vector
 \begin{eqnarray}
 D=
 \begin{pmatrix}
 1\cr
 2
 \end{pmatrix}
 \quad \Rightarrow \quad
 D\,D^\dagger=\begin{pmatrix}
 1 & 2\cr
 2 & 4
 \end{pmatrix}\quad, \quad D^\dagger D=5
 \end{eqnarray}
 Thus, $D D^\dagger$ has eigenvalues $0$ and $5$, while $D^\dagger D$ clearly only has eigenvalue $5$, the difference in rank being accounted for by the count of zero modes.
 
 The zero mode contribution to the propagator can be separated by writing it in one of two ways, depending on which chirality supports zero modes, 
  \bea
  \frac{1}{\slashed{D}+m}&=&
 \begin{pmatrix}
 \frac{m}{m^2+DD^\dagger} &&  \frac{-1}{m^2+DD^\dagger} \, D\cr
 D^\dagger  \frac{1}{m^2+DD^\dagger}&& \left(\frac{1}{m}-\frac{1}{m}D^\dagger  \frac{1}{m^2+DD^\dagger} D\right)
 \end{pmatrix} \nonumber\\
 &=&
 \begin{pmatrix}
\left(\frac{1}{m}-\frac{1}{m}D  \frac{1}{m^2+D^\dagger D} D^\dagger\right) &&  -D\frac{1}{m^2+D^\dagger D} \cr
   \frac{1}{m^2+D^\dagger D}D^\dagger && \frac{m}{m^2+D^\dagger D}
 \end{pmatrix}
 \eea

 An important set of quark bilinears involve the spin tensor $\Sigma_{\mu\nu}$:
 \bea
 \Sigma_{\mu\nu}&=&\frac{1}{2i}[\gamma_\mu , \gamma_\nu]
=\begin{pmatrix}
\bar\sigma_{\mu\nu} & 0\cr
0 & \sigma_{\mu\nu}
\end{pmatrix}
\eea
This representation makes clear the natural decomposition of  $\Sigma_{\mu\nu}$ into its self-dual part ($\sigma_{\mu\nu}$) and its anti-self-dual part ($\bar\sigma_{\mu\nu}$). The bilinears are 
\bea
< \bar\psi \Sigma_{\mu\nu}\psi> ={\rm tr}\left(\Sigma_{\mu\nu}  \frac{1}{\slashed{D}+m}\right)
\label{bilinear}
\eea
using the $2\times 2$ sub-block structure of the propagator, it is straightforward to derive the relations:
\begin{eqnarray}
< \bar\psi \Sigma_{\mu\nu}\psi> &=&m\, {\rm tr}_{2\times 2}
\left(\sigma^{\mu\nu}\frac{1}{m^2+DD^\dagger}+\bar{\sigma}^{\mu\nu}\frac{1}{m^2+ D^\dagger\, D}\right)\\
< \bar\psi \psi> &=& m\, {\rm tr}_{2\times 2}
\left(\frac{1}{m^2+D\, D^\dagger}+\frac{1}{m^2+ D^\dagger\, D}\right)\\
< \bar\psi \Sigma_{\mu\nu}\gamma_5\psi> &=& m\, {\rm tr}_{2\times 2}
\left(\sigma^{\mu\nu}\frac{1}{m^2+D\, D^\dagger}-\bar{\sigma}^{\mu\nu}\frac{1}{m^2+ D^\dagger\, D}\right)\\
< \bar\psi \gamma_5 \psi> &=& m\, {\rm tr}_{2\times 2}
\left(\frac{1}{m^2+D\, D^\dagger}-\frac{1}{m^2+ D^\dagger\, D}\right)
\label{matrix}
\end{eqnarray}
For applications to the chiral magnetic effect, we are interested in the magnetic and electric dipole moments:
\bea
\sigma^{M}_i&=&\frac{1}{2}\epsilon_{ijk}< \bar\psi \Sigma_{jk}\psi> 
\eea
\bea\label{dip123}
\sigma^{E}_i&=&< \bar\psi \Sigma_{i4}\psi>
\eea
With a strong magnetic field in the $x_3$ direction, we concentrate on $\sigma^{M}_3$ and $\sigma^{E}_3$, which require the spin tensors:
\bea\label{eldip2}
\Sigma_{12}
= \begin{pmatrix}
\sigma_{3} & 0\cr
0 & \sigma_{3}
\end{pmatrix}\qquad, \qquad
\Sigma_{34}
=\begin{pmatrix}
-\sigma_{3} & 0\cr
0 & \sigma_{3}
\end{pmatrix}
\eea
Thus,
\bea
m< \bar\psi \Sigma_{12}\psi> &=& {\rm tr}_{2\times 2}\left(\sigma_{3}   \frac{m^2}{m^2+DD^\dagger}\right)+{\rm tr}_{2\times 2}\left(\sigma_{3}   \frac{m^2}{m^2+D^\dagger D}\right)\\
m < \bar\psi \Sigma_{34}\psi> &=&-{\rm tr}_{2\times 2}\left(\sigma_{3}   \frac{m^2}{m^2+DD^\dagger}\right)+{\rm tr}_{2\times 2}\left(\sigma_{3}   \frac{m^2}{m^2+D^\dagger D}\right)
\label{bilinear2}
\eea
The dominant contribution to the trace over the spectrum comes from the modes with low eigenvalues of $DD^\dagger$ and $D^\dagger D$. 
In the strong magnetic field limit,  the zero modes and the near-zero-modes all have spin up, $\sigma_3=+1$, as expected. The dominant contribution to the electric and magnetic moments are therefore:
\bea
m< \bar\psi \Sigma_{12}\psi> &\approx &{\rm tr}_{2\times 2}\left(\frac{m^2}{m^2+DD^\dagger}\right)+{\rm tr}_{2\times 2}\left(\frac{m^2}{m^2+D^\dagger D}\right)\\
m < \bar\psi \Sigma_{34}\psi> &\approx&-{\rm tr}_{2\times 2}\left( \frac{m^2}{m^2+DD^\dagger}\right)+{\rm tr}_{2\times 2}\left(   \frac{m^2}{m^2+D^\dagger D}\right)
\label{bilinear3}
\eea

In particular, this means that in the strong magnetic field limit, we expect
\begin{eqnarray}
\frac{< \bar\psi \Sigma_{12}\psi>}{< \bar\psi \psi>}\to 1\qquad, \qquad B\to\infty
\label{comp1}
\end{eqnarray}
as has been confirmed in a lattice study \cite{Buividovich:2009wi}.

For the magnetic dipole moment, the main contribution comes from the zero modes, so we simply count the degeneracies in the various sectors:
\bea
m < \bar\psi \Sigma_{12}\psi> &\approx&\left(\frac{B-F}{2\pi}\right)\left(\frac{F}{2\pi}\right)
+\left(\frac{B+F}{2\pi}\right)\left(\frac{F}{2\pi}\right)\nonumber\\
&=&\frac{BF}{2\pi^2}
\label{magnetic}
\eea
which is linear in the magnetic field $B$.
For the electric dipole moment, the near-zero-modes cancel, leaving just  the zero mode contribution:
\bea
m < \bar\psi \Sigma_{34}\psi> &\approx&-\left(\frac{B-F}{2\pi}\right)\left(\frac{F}{2\pi}\right)
+\left(\frac{B+F}{2\pi}\right)\left(\frac{F}{2\pi}\right)\nonumber\\
&=&\frac{F^2}{2\pi^2}
\label{electric}
\eea
which is independent of $B$, and negligible compared to $BF$, for $B\gg F$. [Note that (\ref{electric}) does not imply that there is a residual electric dipole moment when $B$ vanishes, because (\ref{electric}) applies only in the $B\gg F$ limit.]
Thus, we see that the zero modes and near-zero-modes imply that
\bea
< \bar\psi \Sigma_{12}\psi>\,\, \propto\,\, B\qquad , \qquad < \bar\psi \Sigma_{12}\psi> \,\,\gg\,\, < \bar\psi \Sigma_{34}\psi>
\eea
This is in agreement with the lattice results of \cite{Buividovich:2009my}.  We note that in a full QCD calculation with dynamical quarks there is an additional instanton measure factor that scales as $m^{N_f}$, which should be take into account for these matrix elements.

If we now consider the fluctuations in the electric dipole moment, we find a dependence on $B$, because
\bea
< \bar\psi \Sigma_{34}\psi\,  \bar\psi \Sigma_{34}\psi>&=&
{\rm tr}\left(  \frac{1}{\slashed{D}+m}\,\Sigma_{34}\, \frac{1}{\slashed{D}+m}\,\Sigma_{34}\right)\nonumber\\
&\hskip -2 cm = & \hskip -1cm{\rm tr}_{2\times 2}\left(\frac{m^2}{(m^2+DD^\dagger)^2}+\frac{1}{(m^2+DD^\dagger)}\,D\sigma_3D^\dagger\sigma_3 \frac{1}{(m^2+DD^\dagger)}\right)+\nonumber\\
& & \hskip -2 cm{\rm tr}_{2\times 2}\left(\frac{1}{(m^2+D^\dagger D)^2}D^\dagger\sigma_3D\sigma_3+\frac{m^2}{(m^2+D^\dagger D)}\sigma_3  \frac{1}{(m^2+D^\dagger D)}\sigma_3\right)\nonumber\\
&\approx& {\rm tr}_{2\times 2}\left(\frac{1}{(m^2+DD^\dagger)}+\frac{1}{(m^2+D^\dagger D)}\right)
\eea
where in the last step we have used the fact that the dominant contribution comes from zero modes and near-zero-modes, all of which have $\sigma_3=+1$. Thus, comparing with (\ref{magnetic}) we see that the fluctuation is linear in $B$
\bea
< \bar\psi \Sigma_{34}\psi\,  \bar\psi \Sigma_{34}\psi> \approx \left(\frac{ F}{2\pi^2 m^2 L^4}\right)\, B
\label{electric-fluc}
\eea
again in agreement with the lattice results of \cite{Buividovich:2009my}.

\section{Conclusions}

In this article we have presented several different perspectives of the chiral magnetic effect. The unifying theme is that the effect arises due to the spin projection of the lowest-Landau-level projection that occurs in a very strong magnetic field, in conjunction with a chirality projection that relates to the axial anomaly, or to the effect of a topologically non-trivial background field such as an instanton. The effect could occur in an abelian theory with an electric field parallel to the strong magnetic field. In this case, the roles of the charge and chiral chemical potentials are played by $A_0$ and $A_3$, respectively, and the chiral magnetic effect is seen to be precisely equivalent to the 2d axial anomaly, which is itself the dimensionally reduced LLL projection of the 4d axial anomaly. Taking this dimensional reduction seriously, in a  theory with a continuous chiral symmetry, we learn further from the 2d physics that there is a spiral condensate, an immediate consequence of the relativistic form of the Peierls instability. Finally, we considered the effect on light quarks of both a magnetic field and an instanton field, showing that the competition between the chiral projection in the instanton field and the spin projection in the magnetic field is responsible for the chiral magnetic effect. We demonstrated that this is consistent with the index theorem, and illustrated the mechanism with a soluble model by taking the leading derivative expansion form of the instanton field in which the magnetic length is much smaller than the instanton scale. It would be interesting to investigate the systematic corrections to this leading approximation using the Fock-Schwinger gauge \cite{Shifman:1980ui,Dubovikov:1981bf,Ioffe:1983ju} representation of the instanton field.

\bigskip
We thank T. Blum, D. Kharzeev and H-U. Yee for helpful discussions.
This work was supported by the US Department of Energy under grants DE-FG02-92ER40716 (GD) and DE-AC02-98CH10886, DE-FG-88ER41723 (GB).


\begin{thebibliography}{999}

\bibitem{Kharzeev:2004ey}
  D.~E.~Kharzeev,
 ``Parity violation in hot QCD: Why it can happen, and how to look for it,''
  Phys.\ Lett.\  B {\bf 633}, 260 (2006)
  [\hhref{hep-ph/0406125}].

\bibitem{arXiv:0706.1026} 
  D.~Kharzeev and A.~Zhitnitsky,
  ``Charge separation induced by P-odd bubbles in QCD matter,''
  Nucl.\ Phys.\ A\ {\bf 797}, 67  (2007)
  [\hhref{0706.1026}[hep-ph]].
  
\bibitem{arXiv:0711.0950} 
  D.~E.~Kharzeev, L.~D.~McLerran and H.~J.~Warringa,
  ``The Effects of topological charge change in heavy ion collisions: 'Event by event P and CP violation',''
  Nucl.\ Phys.\ A\ {\bf 803}, 227  (2008)
  [\hhref{0711.0950}[hep-ph]].
  
  \bibitem{Fukushima:2008xe}
  K.~Fukushima, D.~E.~Kharzeev and H.~J.~Warringa,
  ``The Chiral Magnetic Effect,''
  Phys.\ Rev.\  D {\bf 78}, 074033 (2008)
  [\hhref{0808.3382}[hep-ph]].
  
  \bibitem{Kharzeev:2009fn}
  D.~E.~Kharzeev,
  ``Topologically induced local P and CP violation in QCD x QED,''
  Annals Phys.\  {\bf 325}, 205 (2010)
  [\hhref{0911.3715}[hep-ph]].
  
  \bibitem{Warringa:2012bq} 
  H.~J.~Warringa,
  ``Dynamics of the Chiral Magnetic Effect in a weak magnetic field,''
  arXiv:1205.5679 [hep-th].

\bibitem{Jackiw:1983nv}
  R.~Jackiw,
  ``Topological Investigations Of Quantized Gauge Theories,'' in S. B. Treiman, R. Jackiw, B. Zumino and E. Witten,  {\it Current Algebra and Anomalies} (Princeton University Press, 1985).
  
 \bibitem{Shifman:1999mk}
  M.~A.~Shifman,
  {\it ITEP lectures on particle physics and field theory},  Vol. 1, 2,
  World Sci.\ Lect.\ Notes Phys.\  {\bf 62},  (1999).
  
  \bibitem{Gusynin:1995nb}
  V.~P.~Gusynin, V.~A.~Miransky and I.~A.~Shovkovy,
 ``Dimensional reduction and catalysis of dynamical symmetry breaking by a
magnetic field,''
  Nucl.\ Phys.\  B {\bf 462}, 249 (1996)
  [\hhref{hep-ph/9509320}].
  
  \bibitem{strang}
  G. Strang, {\it Linear algebra and its applications},  (Harcourt, San Diego, 1988).

  \bibitem{aharonov-casher}
  Y. Aharonov and A. Casher 
  ``Ground state of a spin-1/2 charged particle in a two-dimensional magnetic field, ''
  Phys.\ Rev.\ A {\bf 19}, 2461 (1979).
  
    \bibitem{novikov}
S.~P.~Novikov and B.~A.~Dubrovin, 
   ``Ground states of a two-dimensional electron in a periodic magnetic field, '' 
   Zh.\ Eksper.\ Teoret.\ Fiz.\ , {\bf 79}, 1006 (1980), 
 [Sov. Phys. JETP {\bf 52}, 511 (1980)],
 ``Ground states in a periodic field. Magnetic Bloch functions and vector bundles,'' 
 Dokl.\ Akad.\ Nauk SSSR,\ {\bf 253}, 1293 (1980).

\bibitem{Brown:1977bj}
  L.~S.~Brown, R.~D.~Carlitz, C.~Lee,
  ``Massless Excitations in Instanton Fields,''
  Phys.\ Rev.\  {\bf D16}, 417 (1977);
  R.~D.~Carlitz, C.~Lee,
  ``Physical Processes In Pseudoparticle Fields: The Role Of Fermionic Zero Modes,''
  Phys.\ Rev.\  {\bf D17}, 3238 (1978).

\bibitem{Hur:2010bd} 
  J.~Hur, C.~Lee and H.~Min,
  ``Some chirality-related properties of the 4-D massive Dirac propagator and determinant in an arbitrary gauge field,''
  Phys.\ Rev.\ D {\bf 82}, 085002 (2010)
  [arXiv:1007.4616 [hep-th]].

\bibitem{bardeen}
 W.~A.~Bardeen and B.~Zumino,
 ``Consistent and Covariant Anomalies in Gauge and Gravitational
Theories,''
 Nucl.\ Phys.\ B {\bf 244}, 421 (1984).

\bibitem{dt}
 G.~V.~Dunne and C.~A.~Trugenberger,
 ``Odd Dimensional Gauge Theories And Current Algebra,''
 Annals Phys.\  {\bf 204}, 281 (1990).
 
 \bibitem{Gorbar:2010kc} 
  E.~V.~Gorbar, V.~A.~Miransky and I.~A.~Shovkovy,
  ``Chiral asymmetry and axial anomaly in magnetized relativistic matter,''
  Phys.\ Lett.\ B {\bf 695}, 354 (2011)
  [arXiv:1009.1656 [hep-ph]].

\bibitem{he}
W. Heisenberg and H. Euler,
``Consequences of Dirac's Theory of Positrons'',
Zeit. f. Phys. {\bf 98}, 714 (1936).
  
\bibitem{schwinger1}
J.~Schwinger,
``On gauge invariance and vacuum polarization'',
Phys. Rev. {\bf 82} , 664 (1951).

  \bibitem{Kluger:1998bm} 
  Y.~Kluger, E.~Mottola and J.~M.~Eisenberg,
  ``The Quantum Vlasov equation and its Markov limit,''
  Phys.\ Rev.\ D {\bf 58}, 125015 (1998)
  [hep-ph/9803372].
  
  \bibitem{Schwinger:1962tp} 
  J.~S.~Schwinger,
  ``Gauge Invariance and Mass,''
  Phys.\ Rev.\  {\bf 125}, 397 (1962);
  ``Gauge Invariance and Mass. 2.,''
  Phys.\ Rev.\  {\bf 128}, 2425 (1962).

\bibitem{Basar:2010zd} 
  G.~Basar, G.~V.~Dunne and D.~E.~Kharzeev,
  ``Chiral Magnetic Spiral,''
  Phys.\ Rev.\ Lett.\  {\bf 104}, 232301 (2010)
  [arXiv:1003.3464 [hep-ph]].
  
\bibitem{Gross:1974jv} 
  D.~J.~Gross and A.~Neveu,
  ``Dynamical Symmetry Breaking in Asymptotically Free Field Theories,''
  Phys.\ Rev.\ D {\bf 10}, 3235 (1974).
  
  \bibitem{quarkyonic}
  T.~Kojo, Y.~Hidaka, L.~McLerran, R.~D.~Pisarski,
  ``Quarkyonic Chiral Spirals,''
  Nucl.\ Phys.\  {\bf A843}, 37-58 (2010).
  [arXiv:0912.3800 [hep-ph]].


  \bibitem{peierls}
R.~Peierls, {\it The Quantum Theory of Solids} (Oxford, 1955), {\it More Surprises in Theoretical Physics} (Princeton, 1991)


\bibitem{basar}
  G.~Basar, G.~V.~Dunne and M.~Thies,
  ``Inhomogeneous Condensates in the Thermodynamics of the Chiral ${\rm NJL}_2$
  model,''
  Phys.\ Rev.\  D {\bf 79}, 105012 (2009)
  [arXiv:0903.1868 [hep-th]];
  G.~Basar and G.~V.~Dunne,
  ``A Twisted Kink Crystal in the Chiral Gross-Neveu model,''
  Phys.\ Rev.\  D {\bf 78}, 065022 (2008)
  [arXiv:0806.2659 [hep-th]].

  \bibitem{schon}
  V.~Schon and M.~Thies,
  ``Emergence of Skyrme crystal in Gross-Neveu and 't Hooft models at  finite density,''
  Phys.\ Rev.\  D {\bf 62}, 096002 (2000)
  [arXiv:hep-th/0003195],
``2D model field theories at finite temperature and density,'' 
in {\it At the frontier of particle physics : handbook of QCD}, Vol. 3, M.~A.~Shifman (ed.) (World Scientific, 2000),  [arXiv:hep-th/0008175].
  
    \bibitem{Thies:2006ti} 
  M.~Thies,
  ``From relativistic quantum fields to condensed matter and back again: Updating the Gross-Neveu phase diagram,''
  J.\ Phys.\ A {\bf 39}, 12707 (2006)
  [hep-th/0601049].
  
  \bibitem{Bzdak:2009fc}
  A.~Bzdak, V.~Koch,  J.~Liao,
  ``Remarks on possible local parity violation in heavy ion collisions,''
  Phys.\ Rev.\ C {\bf 81}, 031901 (2010)
  [arXiv:0912.5050 [nucl-th]].
  
\bibitem{:2009uh}
  B.~I.~Abelev {\it et al.}  [STAR Collaboration],
  ``Azimuthal Charged-Particle Correlations and Possible Local Strong Parity
  Violation,''
  Phys.\ Rev.\ Lett.\  {\bf 103}, 251601 (2009)
  [arXiv:0909.1739 [nucl-ex]].

\bibitem{:2009txa}
  B.~I.~Abelev {\it et al.}  [STAR Collaboration],
  ``Observation of charge-dependent azimuthal correlations and possible local
  strong parity violation in heavy ion collisions,''
  Phys.\ Rev.\ C {\bf 81}, 054908 (2010)
  [arXiv:0909.1717 [nucl-ex]].
    
\bibitem{Kharzeev:1998kz}
  D.~Kharzeev, R.~D.~Pisarski and M.~H.~G.~Tytgat,
  ``Possibility of spontaneous parity violation in hot {QCD},''
  Phys.\ Rev.\ Lett.\  {\bf 81}, 512 (1998)
  [arXiv:hep-ph/9804221].

  \bibitem{holographiccms}
  K.~-Y.~Kim, B.~Sahoo, H.~-U.~Yee,
  ``Holographic chiral magnetic spiral,''
  JHEP {\bf 1010}, 005 (2010).
  [arXiv:1007.1985 [hep-th]].
  
\bibitem{Sakai:2004cn} 
  T.~Sakai and S.~Sugimoto,
  ``Low energy hadron physics in holographic QCD,''
  Prog.\ Theor.\ Phys.\  {\bf 113}, 843 (2005)
  [hep-th/0412141];
  ``More on a holographic dual of QCD,''
  Prog.\ Theor.\ Phys.\  {\bf 114}, 1083 (2005)
  [hep-th/0507073].
  
\bibitem{Basar:2011by} 
  G.~Basar, G.~V.~Dunne and D.~E.~Kharzeev,
  ``Electric dipole moment induced by a QCD instanton in an external magnetic field,''
  Phys.\ Rev.\ D {\bf 85}, 045026 (2012)
  [arXiv:1112.0532 [hep-th]].

\bibitem{arXiv:0907.1396} 
  V.~Skokov, A.~Y.~Illarionov and V.~Toneev,
  ``Estimate of the magnetic field strength in heavy-ion collisions,''
  Int.\ J.\ Mod.\ Phys.\ A\ {\bf 24}, 5925  (2009)
  [\hhref{0907.1396}[nucl-th]].

\bibitem{arXiv:1111.1949} 
  A.~Bzdak and V.~Skokov,
  ``Event-by-event fluctuations of magnetic and electric fields in heavy ion collisions,''
  Phys.\ Lett.\ B {\bf 710}, 171 (2012)
  [arXiv:1111.1949 [hep-ph]].
  
   \bibitem{Buividovich:2009wi}
 P.~V.~Buividovich, M.~N.~Chernodub, E.~V.~Luschevskaya, M.~I.~Polikarpov,
  ``Numerical study of chiral symmetry breaking in non-Abelian gauge theory with background magnetic field,''
  Phys.\ Lett.\  {\bf B682}, 484-489 (2010)
  [\hhref{0812.1740} [hep-lat]];
   ``Chiral magnetization of non-Abelian vacuum: A Lattice study,''
  Nucl.\ Phys.\  {\bf B826}, 313-327 (2010)
  [\hhref{0906.0488} [hep-lat]];
  ``Numerical evidence of chiral magnetic effect in lattice gauge theory,''
  Phys.\ Rev.\  D {\bf 80}, 054503 (2009)
  [\hhref{0907.0494} [hep-lat]].
  
\bibitem{Buividovich:2009my}
  P.~V.~Buividovich, M.~N.~Chernodub, E.~V.~Luschevskaya, M.~I.~Polikarpov,
  ``Quark electric dipole moment induced by magnetic field,''
  Phys.\ Rev.\  {\bf D81}, 036007 (2010).
  [\hhref{0909.2350} [hep-ph]].

  
  \bibitem{Abramczyk:2009gb}
  M.~Abramczyk, T.~Blum, G.~Petropoulos and R.~Zhou,
  ``Chiral magnetic effect in 2+1 flavor QCD+QED,''
 PoS {\bf LAT2009}, 181 (2009).
  [\hhref{0911.1348} [hep-lat]].
  
  \bibitem{tom}
  T.~Blum, talk at Workshop on
P- and CP-odd Effects in Hot and Dense Matter, Brookhaven National Laboratory, April, 2010.

\bibitem{Braguta:2010ej}
  V.~V.~Braguta, P.~V.~Buividovich, T.~Kalaydzhyan, S.~V.~Kuznetsov, M.~I.~Polikarpov,
  ``The Chiral Magnetic Effect and chiral symmetry breaking in SU(3) quenched lattice gauge theory,''
  PoS {\bf LATTICE2010}, 190 (2010).
  [\hhref{1011.3795} [hep-lat]].

  \bibitem{Tiburzi:2011vk} 
  B.~C.~Tiburzi,
  ``Lattice QCD with Classical and Quantum Electrodynamics,''
  PoS LATTICE {\bf 2011}, 020 (2011)
  [arXiv:1110.6842 [hep-lat]].
     
   \bibitem{Giusti:2001ta}
  L.~Giusti, A.~Gonzalez-Arroyo, C.~Hoelbling, H.~Neuberger and C.~Rebbi,
  ``Fermions on tori in uniform Abelian fields,''
  Phys.\ Rev.\  D {\bf 65}, 074506 (2002)
 [\hhref{hep-lat/0112017}].
  
 \bibitem{Tenjinbayashi:2005sy}
  Y.~Tenjinbayashi, H.~Igarashi and T.~Fujiwara,
  ``Dirac operator zero-modes on a torus,''
  Annals Phys.\  {\bf 322}, 460 (2007)
  [\hhref{hep-th/0506259}].
  
   \bibitem{AlHashimi:2008hr}
  M.~H.~Al-Hashimi and U.~J.~Wiese,
  ``Discrete Accidental Symmetry for a Particle in a Constant Magnetic Field on a Torus,''
  Annals Phys.\  {\bf 324}, 343 (2009)
  [\hhref{0807.0630}[quant-ph]].
  
  \bibitem{Zak:1964zz}
  J.~Zak,  ``Magnetic Translation Group,''
  Phys.\ Rev.\  {\bf 134}, A1602 (1964). 
  
  \bibitem{'tHooft:1976fv}
  G.~'t Hooft,
  ``Computation of the Quantum Effects Due to a Four-Dimensional Pseudoparticle,''
  Phys.\ Rev.\  {\bf D14}, 3432 (1976).
  
\bibitem{Schwarz:1977az}
  A.~S.~Schwarz,
  ``On Regular Solutions Of Euclidean Yang-Mills Equations,''
  Phys.\ Lett.\  B {\bf 67}, 172 (1977).
  
\bibitem{Kiskis:1977vh}
J.~E.~Kiskis,
``Fermions In A Pseudoparticle Field,''
 Phys.\ Rev.\  D {\bf 15}, 2329 (1977).
  
\bibitem{Jackiw:1977pu}
  R.~Jackiw and C.~Rebbi,
  ``Spinor analysis of Yang-Mills theory,''
  Phys.\ Rev.\  D {\bf 16}, 1052 (1977).
  
   \bibitem{rubakov}
  V.~A.~Rubakov, 
  {\it Classical Theory of Gauge Fields},
  (Princeton Univ. Press, 2002).

  
\bibitem{Jackiw:1976dw}
  R.~Jackiw and C.~Rebbi,
  ``Conformal properties of a Yang-Mills pseudoparticle,''
  Phys.\ Rev.\  D {\bf 14}, 517 (1976).
  
\bibitem{Chadha:1977mh}
  S.~Chadha, A.~D'Adda, P.~Di Vecchia and F.~Nicodemi,
  ``Fermions In The Background Pseudoparticle Field In An O(5) Formulation,''
  Phys.\ Lett.\  B {\bf 67}, 103 (1977).

\bibitem{Belavin:1975fg}
  A.~A.~Belavin, A.~M.~Polyakov, A.~S.~Schwartz and Yu.~S.~Tyupkin,
  ``Pseudoparticle solutions of the Yang-Mills equations,''
  Phys.\ Lett.\  B {\bf 59}, 85 (1975).
  
  \bibitem{aps}
  M. Atiyah, V. Patodi, and I. Singer, 
  ``Spectral asymmetry and Riemannian geometry'', 
  Math. Proc. Camb. Philos. Soc. {\bf 77}, 43 (1975).

\bibitem{'tHooft:1981sz}
  G.~'t Hooft,
  ``Some Twisted Selfdual Solutions For The Yang-Mills Equations On A
  Hypertorus,''
  Commun.\ Math.\ Phys.\  {\bf 81}, 267 (1981).

\bibitem{vanBaal}
P.~van Baal,
  ``Some Results For SU(N) Gauge Fields On The Hypertorus,''
  Commun.\ Math.\ Phys.\  {\bf 85}, 529 (1982),
  ``SU(N) Yang-Mills Solutions With Constant Field Strength On T4,''
  Commun.\ Math.\ Phys.\  {\bf 94}, 397 (1984),
   P.~van Baal,
  ``Instanton moduli for T(3)xR,''
  Nucl.\ Phys.\ Proc.\ Suppl.\  {\bf 49}, 238 (1996)
  [\hhref{hep-th/9512223}].
  
\bibitem{Shifman:1980ui}
  M.~A.~Shifman,
  ``Wilson Loop in Vacuum Fields,''
  Nucl.\ Phys.\  {\bf B173}, 13 (1980).
  
\bibitem{Dubovikov:1981bf}
  M.~S.~Dubovikov, A.~V.~Smilga,
  ``Analytical Properties of the Quark Polarization Operator in an External Selfdual Field,''
  Nucl.\ Phys.\  {\bf B185}, 109-132 (1981).

\bibitem{Ioffe:1983ju}
  B.~L.~Ioffe, A.~V.~Smilga,
  ``Nucleon Magnetic Moments and Magnetic Properties of Vacuum in QCD,''
  Nucl.\ Phys.\  {\bf B232}, 109 (1984).
 
\end{thebibliography}
\end{document}